\documentclass[aps,prl,showpacs,english,10pt,a4,nofootinbib,notitlepage,twocolumn,superscriptaddress]{revtex4-1}

\usepackage[utf8]{inputenc}
\usepackage[T1]{fontenc}
\usepackage[english]{babel}  

\usepackage{times}
\usepackage{dsfont}

\usepackage{amssymb,amsmath,amsfonts,amsthm}
\usepackage{mathtools}
\usepackage{verbatim}
\usepackage{enumerate}
\usepackage{graphicx}
\graphicspath{{figs/}}
\usepackage{hyperref}
\usepackage{bbm}
\usepackage{booktabs}

\usepackage[caption=false]{subfig}

\usepackage{color}
\definecolor{dred}{rgb}{.8,0.2,.2}
\definecolor{ddred}{rgb}{.8,0.5,.5}
\definecolor{dblue}{rgb}{.2,0.2,.8}
\definecolor{dgreen}{rgb}{.2,0.5,.2}

\newcommand{\bra}[1]{\mbox{$\langle #1|$}}
\newcommand{\ket}[1]{\ensuremath{|#1\rangle}}

\newcommand{\ketbra}[2]{\mbox{$|#1\rangle\langle #2|$}}
\newcommand{\tr}{\textrm{tr}}

\newcommand{\be}{\begin{equation}}
\newcommand{\ee}{\end{equation}}

\newcommand{\bea}{\begin{eqnarray}}
\newcommand{\eea}{\end{eqnarray}}

\begin{document}

\title{Quantum state tomography via reduced density matrices}

\author{Tao Xin}
\thanks{These authors contributed equally to this work.}
\affiliation{State Key Laboratory of Low-Dimensional Quantum Physics and Department of Physics, Tsinghua University, Beijing 100084, China}
\affiliation{Institute for Quantum Computing,
University of Waterloo, Waterloo N2L 3G1, Ontario, Canada}

\author{Dawei Lu}
\thanks{These authors contributed equally to this work.}
\affiliation{Institute for Quantum Computing,
University of Waterloo, Waterloo N2L 3G1, Ontario, Canada}
\affiliation{Department of Physics and Astronomy, University of Waterloo, Waterloo, Ontario N2L 3G1, Canada}

\author{Joel Klassen}
\thanks{These authors contributed equally to this work.}
\affiliation{Institute for Quantum Computing,
University of Waterloo, Waterloo N2L 3G1, Ontario, Canada}
\affiliation{Department of Physics, University of
  Guelph, Guelph, Ontario, Canada}

\author{Nengkun Yu}
\email{nengkunyu@gmail.com}
\affiliation{Institute for Quantum Computing,
University of Waterloo, Waterloo N2L 3G1, Ontario, Canada}
\affiliation{Centre for Quantum Computation \& Intelligent Systems,
   Faculty of Engineering and Information Technology, University of
   Technology Sydney, NSW 2007, Australia}
\affiliation{Department of Mathematics \& Statistics, University of
  Guelph, Guelph, Ontario, Canada}%

\author{Zhengfeng Ji}
\affiliation{Centre for Quantum Computation \& Intelligent Systems,
   Faculty of Engineering and Information Technology, University of
   Technology Sydney, NSW 2007, Australia}
\affiliation{State Key Laboratory of Computer Science, Institute of
  Software, Chinese Academy of Sciences, Beijing, China}%

\author{Jianxin Chen}
\address{Joint Center for Quantum Information and Computer Science,
  University of Maryland, College Park, Maryland, USA}

\author{Xian Ma}
\affiliation{Institute for Quantum Computing,
University of Waterloo, Waterloo N2L 3G1, Ontario, Canada}
\affiliation{Department of Physics and Astronomy, University of Waterloo, Waterloo, Ontario N2L 3G1, Canada}

\author{Guilu Long}
\affiliation{State Key Laboratory of Low-Dimensional Quantum Physics and Department of Physics, Tsinghua University, Beijing 100084, China}

\author{Bei Zeng}
\email{zengb@uoguelph.ca}
\affiliation{Institute for Quantum Computing,
University of Waterloo, Waterloo N2L 3G1, Ontario, Canada}
\affiliation{Department of Mathematics \& Statistics, University of
  Guelph, Guelph, Ontario, Canada}
\affiliation{Canadian Institute for Advanced Research, Toronto,
  Ontario, Canada}

\author{Raymond Laflamme}
\affiliation{Institute for Quantum Computing,
University of Waterloo, Waterloo N2L 3G1, Ontario, Canada}
\affiliation{Department of Physics and Astronomy, University of Waterloo, Waterloo, Ontario N2L 3G1, Canada}
\affiliation{Canadian Institute for Advanced Research, Toronto,
  Ontario, Canada}
\affiliation{Perimeter Institute for Theoretical Physics, Waterloo N2L 2Y5, Ontario,
Canada}%

\begin{abstract}
Quantum state tomography via local measurements is an efficient tool for characterizing quantum states. However it requires that the original global state be uniquely determined (UD) by its local reduced density matrices (RDMs). In this work we demonstrate for the first time a class of states that are UD by their RDMs under the assumption that the global state is pure, but fail to be UD in the absence of that assumption. This discovery allows us to classify quantum states according to their UD properties, with the requirement that each class be treated distinctly in the practice of simplifying quantum state tomography. Additionally we experimentally test the feasibility and stability of performing quantum state tomography via the measurement of local RDMs for each class. These theoretical and experimental results advance the project of performing efficient and accurate quantum state tomography in practice.
\end{abstract}

\maketitle

\textit{Introduction}---Quantum state tomography (QST) is one of the most famous double-edged swords in quantum information science. On the one hand, QST provides a complete description of an arbitrary quantum state, which is important in benchmarking and validating quantum devices \cite{d2002quantum,haffner2005scalable,leibfried2005creation,lvovsky2009continuous,baur2012benchmarking}. On the other hand, the exponential resources QST requires make scaling it to large systems infeasible in practice. In the past decade, tremendous effort has been devoted to boosting the efficiency of QST \cite{KIRI08,VHCA13,FGLE12,GLFB10,LXYJ15,BDK15,KB15,HHJW16}. Among them, QST via reduced density matrices (RDMs) \cite{Linden2002,Linden2002b,Diosi2004,CJRZ12,Chen2012,Chen2013} has been one especially promising approach, as many experimental setups are able to perform local measurements conveniently and accurately. One criterion for adopting this approach is that the global state has to be the only state which is compatible with its RDMs, that is, it must be uniquely determined (UD) by its RDMs.

The UD criterion can be further classified into two categories: uniquely determined among all states (UDA) and uniquely determined among pure states (UDP) by local RDMs\footnote{In this work UD refers to UD by its RDMs unless otherwise specified. Note there are some other UD sources with different properties, but they are not considered here. See appendix A \cite{supple} for details.}. The ground states of many physically realistic quantum systems usually belong to the UDA category. These systems involve only few-body interactions~\cite{hastings2010locality}, and possess ground states which exhibit special properties~\cite{wolf2008area,perez2006matrix,verstraete2008matrix,cirac2009renormalization}. To reconstruct states of this type, experimentalists need only measure RDMs and search for the global state which is compatible with these RDMs. This saves an exponential number of measurements~\cite{cramer2010efficient}.

In the case of states which satisfy the UDP criterion, two assumptions must be made if one wishes to reconstruct such states via RDMs. First, the experimentally prepared states must be (nearly) pure. Second, the search space of possible reconstructions must be limited to pure states, otherwise the searching procedure may return incorrect mixed states with the same RDMs. Despite these assumptions, searching for UDP states has the advantage of significantly reducing the number of search parameters, since the searching procedure is restricted in the pure state space. Traditionally this has been the approach for dealing with many related problems, for instance, the famous Pauli problem, which asks whether the probability distribution of position and momentum is enough to determine the wave function~\cite{pauli1958allgemeinen,weigert1992pauli}.

Obviously, UDA implies UDP. One notable hypothesis is that UDP also implies UDA \cite{CJRZ12}, deriving from the fact that UDP implies UDA in 3-qubit systems \cite{Linden2002}.
It is then natural to ask whether this hypothesis is true in general. If the answer is \emph{yes}, it would permit experimentalists to preserve the substantially fewer searching parameters even when dealing with the UDA category.

\begin{figure}[h]
\includegraphics[scale=.4]{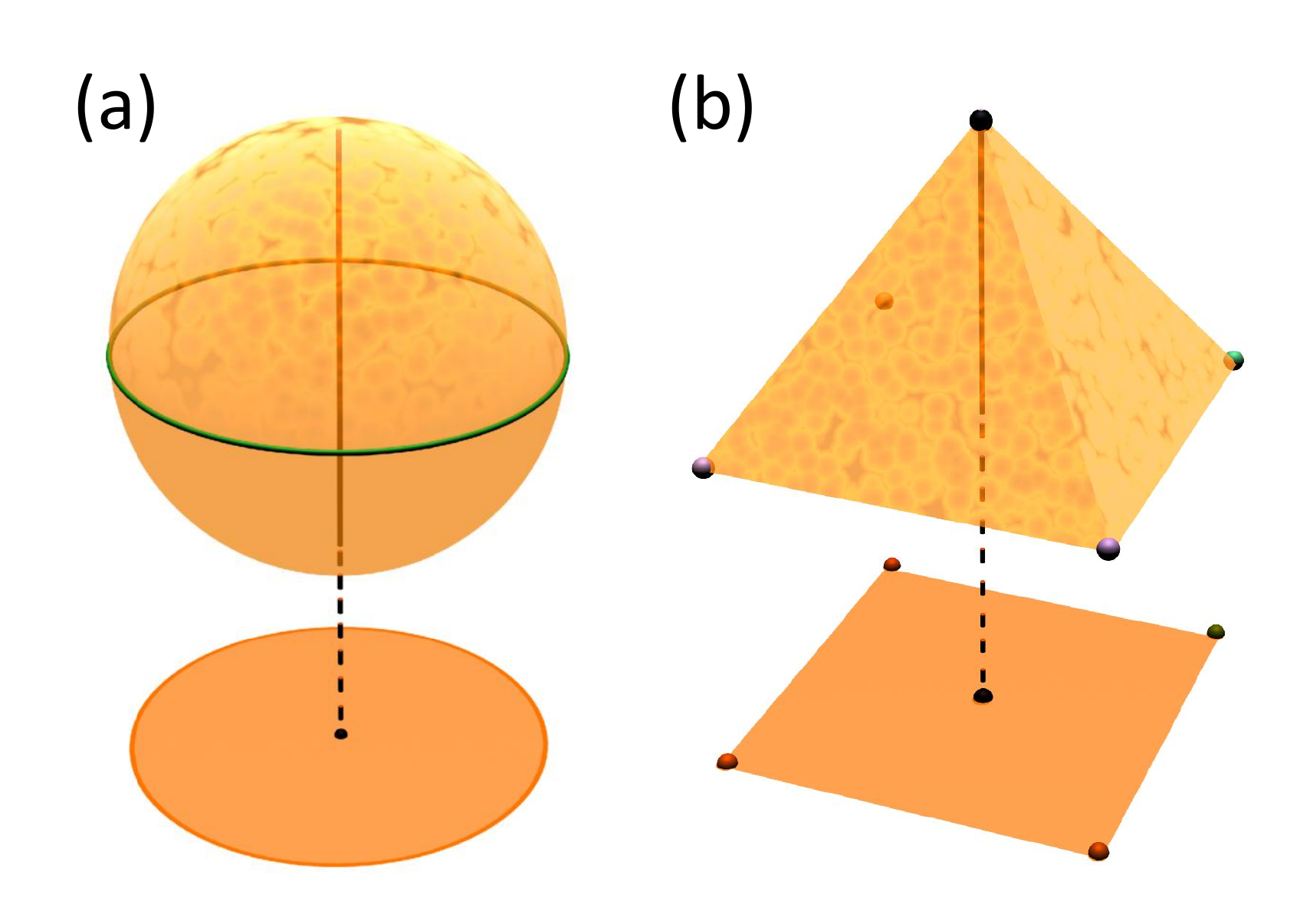}
\caption{Three-dimensional caricatures of the possible shapes of state space, and the space of reduced density matrices as projections. Pure states are given by the extreme points. (a) A sphere, for which all boundary points are extreme points. Only the points on the boundary of the projected circle have a unique preimage in the state space, and so are UDA. All the interior points have multiple extreme points in their preimage, so they are not UDP. Thus UDP implies UDA. (b) A polytope, for which the five vertices are extreme points. The four corner points have a unique preimage in the state space, and so are UDA. However, one interior point located at the centre has  multiple preimages where only one is an extreme point, so it is UDP but not UDA.}
\label{UDAUDP}
\end{figure}

Unfortunately, the validity of this hypothesis becomes a mystery in the study of QST (see appendix A \cite{supple} for more historical researches). To date, no systematic method has been proposed to verify this hypothesis, except concrete examples~\cite{heinosaari2013quantum,Chen2013}.
To comprehensively understand this problem, it is instructive to think about the geometric picture illustrated in Fig.~\ref{UDAUDP}.
Figure~\ref{UDAUDP}(a) is a more familiar shape of state space (e.g. the Bloch sphere) for which UDP implies UDA. However, in higher dimensional state spaces, regions could possibly look like Fig.~\ref{UDAUDP}(b), where some points can be UDP but not UDA.

In this work we disprove the above hypothesis by showing that UDP does not always imply UDA. In particular, we present a class of 4-qubit states that are UDP by their two-particle RDMs ($2$-RDMs), but fail to be UDA. This is the first separation between UDA and UDP in the setting of RDMs. Our construction is based on the studys of 4-qubit symmetric ($i.e.$ bosonic) states. Note that the properties of bosonic states have recently been extensively studied theoretically \cite{ESBL02,BKMG+09,YU14} and experimentally \cite{CBF+13,MZHC15} due to their significant roles in characterizing cold atomic systems. To illustrate the validity of our construction, we experimentally demonstrate the reconstruction of a series of 4-qubit states by measuring their $2$-RDMs. We examine the differences among states that are: A) Neither UDP nor UDA; B) UDP and UDA; C) UDP but not UDA. We test the robustness (stability) against experimental errors of our construction.

\textit{Three classes}---We classify 4-qubit pure states into three classes according to how they are UD by their 2-RDMs, and present some examples for each class.

\textbf{Class A}: neither UDP nor UDA. Consider the GHZ-type state $\alpha\ket{0000}+\beta\ket{1111}$, whose $2$-RDMs are
\begin{equation}
\label{eq:GHZRDM}
|\alpha|^2\ket{00}\bra{00}+|\beta|^2\ket{11}\bra{11}.
\end{equation}
It is not UDP (thus not UDA) since any pure state $\alpha\ket{0000}+e^{i\phi}\beta\ket{1111}$ or mixed state $|\alpha|^2\ket{0000}\bra{0000}+|\beta|^2\ket{1111}\bra{1111}$ has the same $2$-RDMs.
Therefore, to reconstruct 4-qubit GHZ-type states experimentally, it is insufficient to only measure its $2$-RDMs, even if assuming the prepared state is pure.

\textbf{Class B}: UDP and UDA. The W-type state
\begin{equation}
\label{wstate_e}
\ket{\text{W}}=a\ket{0001}+b\ket{0010}+c\ket{0100}+d\ket{1000},
\end{equation}
is known to be UDA~\cite{parashar2009n}, and also UDP.
Unlike the GHZ-type state, to reconstruct the global state, one needs only know its $2$-RDMs.

\textbf{Class C}: UDP but not UDA. Existence of this type of states is the main theoretical results of this paper. Up until now, no such states are known. This is likely due to the fact that analytically determining the uniqueness properties of quantum states is notoriously difficult in general.

The outline of our approach is as follows. We focus on the $4$-qubit bosonic (symmetric) state $\ket{\psi_S}=\sum_{j=0}^4 c_j \ket{w_j}$,
where the normalized state $\ket{w_j}$ is defined to be proportional to $P_{\textrm{sym}}\bigl( \ket{0}^{\otimes j} \otimes \ket{1}^{\otimes 4-j} \bigr)$ with $P_{\textrm{sym}}$ being the projection onto the 4-qubit symmetric subspace.
This symmetry assumption significantly simplifies the analysis since all the $2$-RDMs are the same. To further simplify the analysis, we assume $c_1=c_3=0$ and $c_0$, $c_2$ and $c_4$ are all real:
\begin{equation}
\label{target_state}
\ket{\psi_S}=c_0 \ket{w_0} + c_2 \ket{w_2} +c_4 \ket{w_4}.
\end{equation}

To determine the parameter regions of $c_0,c_2,c_4$ where $\ket{\psi_S}$ is UDP but not UDA, we take three steps:

\emph{Step 1}. First we prove that there is no other pure bosonic state which has the same $2$-RDMs as $\ket{\psi_S}$ when $\ket{\psi_S}$'s 2-RDMs have three distinct non-zero eigenvalues.

\emph{Step 2}. Next we observe that any pure bosonic state which is uniquely determined among all other pure bosonic states is also UDP.

\emph{Step 3}. Finally we provide the region where the $2$-RDMs of $\ket{\psi_S}$ are separable. $\ket{\psi_S}$ is guaranteed not to be UDA in this region. Therefore, within this parameter region, $\ket{\psi_S}$ is UDP but not UDA as long as its 2-RDMs are non-degenerate and not rank one.

We direct the reader to appendix B for steps 1 and 2, and appendix C for step 3 \cite{supple}.

\begin{figure}[htb]
\begin{center}
\includegraphics[width= 1\columnwidth]{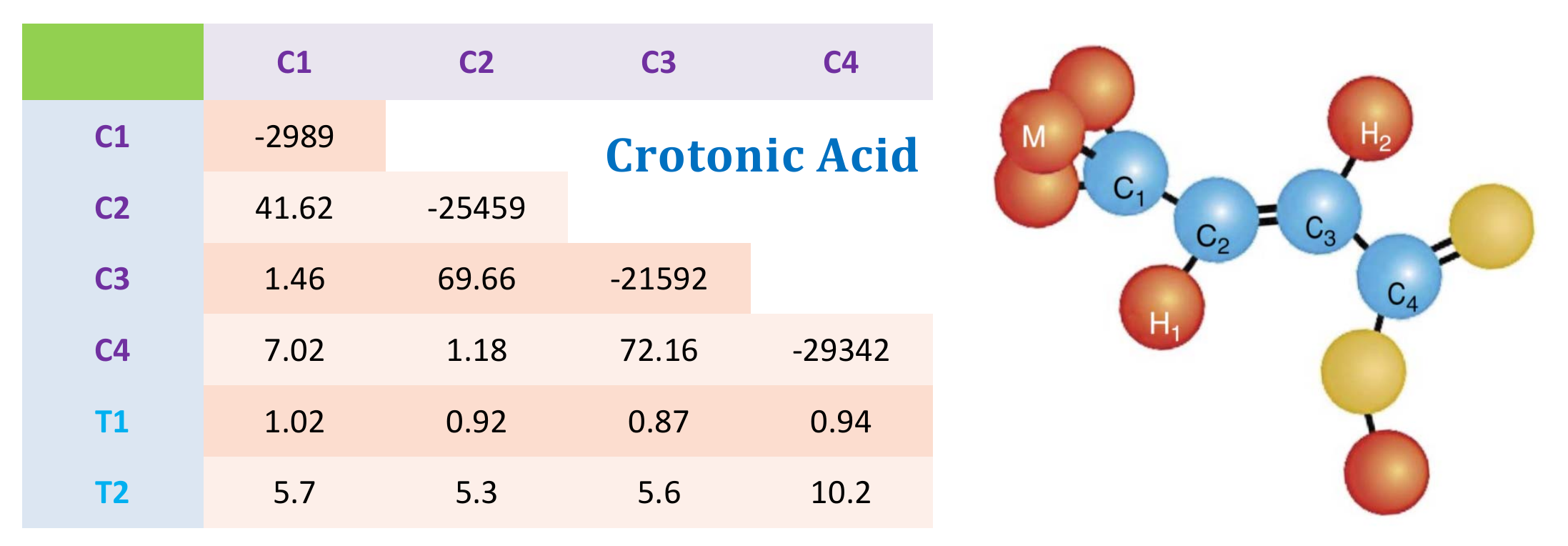}
\end{center}
\setlength{\abovecaptionskip}{-0.00cm}
\caption{\footnotesize{Molecular structure and Hamiltonian parameters of $^{13}$C-labeled trans-crotonic acid. C$_1$, C$_2$, C$_3$ and C$_4$ are used as four qubits. The chemical shifts and J-couplings (in Hz) are listed by the diagonal and off-diagonal elements, respectively. T$_{1}$ and T$_{2}$ (in Seconds) are also shown at bottom.}}\label{molecule}
\end{figure}

\textit{Experiment}---We experimentally inspect all three classes of state using nuclear magnetic resonance (NMR), and test their stability against experimental noise.
The 4-qubit sample is $^{13}$C-labeled trans-crotonic acid dissolved in d6-acetone. The structure of the molecule is shown in Fig. \ref{molecule}, where C$_1$ to C$_4$ denote the four qubits. The methyl group M, H$_1$ and H$_2$ were decoupled throughout all experiments. The internal Hamiltonian under weak coupling approximation is
\begin{align}\label{Hamiltonian}
\mathcal{H}_{\text{int}}=\sum\limits_{j=1}^4 {\pi \nu _j } \sigma_z^j  + \sum\limits_{j < k,=1}^4 {\frac{\pi}{2}} J_{jk} \sigma_z^j \sigma_z^k,
\end{align}
where $\nu_j$ is the chemical shift and $\emph{J}_{jk}$ is the J-coupling strength. All experiments were carried out on a Bruker DRX 700MHz spectrometer at room temperature.

The experiments are divided into three steps: (i) prepare the initial state $\ket{0000}$; (ii) evolve $\ket{0000}$ to the desired state in each class; (iii) measure the final state by full QST and 2-RDMs, reconstruct the original state via the measured 2-RDMs, and compare it with the full QST result. We describe each step briefly as follows. For more experimental details, see appendices E and F \cite{supple}.

(i) In the majority of experiments in quantum information, $\ket{0}^{\otimes n}$ is chosen as the input state. In NMR we instead generate a so-called pseudo-pure state (PPS) from the thermal equilibrium state via the spatial averaging technique~\cite{cory1997ensemble,lu2011simulation,PhysRevA.92.022126}. The form of 4-qubit PPS is
$\rho_{0000}=(1-\epsilon){\mathbb{I}}/16+\epsilon\ket{0000}\bra{0000}$, 
where $\mathbb{I}$ is identity and $\epsilon\approx 10^{-5}$ is the polarization. Only the deviated part $\ket{0000}$ contributes to the NMR signals and the PPS is able to serve as an input state.

(ii) The next step is to create the desired states of the different UD classes. The radio-frequency (RF) pulses during this procedure are optimized by the gradient ascent pulse engineering (GRAPE) algorithm~\cite{khaneja2005optimal,ryan2008liquid}, and are designed to be robust to the static field distributions ($T_2^{*}$ process) and RF inhomogeneity. The designed fidelity for each pulse exceeds 0.99, and all pulses are corrected via a feedback-control setup in the NMR spectrometer to minimize the discrepancies between the ideal and implemented pulses~\cite{moussa2012practical,lu2015experimental}.

\begin{figure}[htb]
\begin{center}
\includegraphics[width= 0.9\columnwidth]{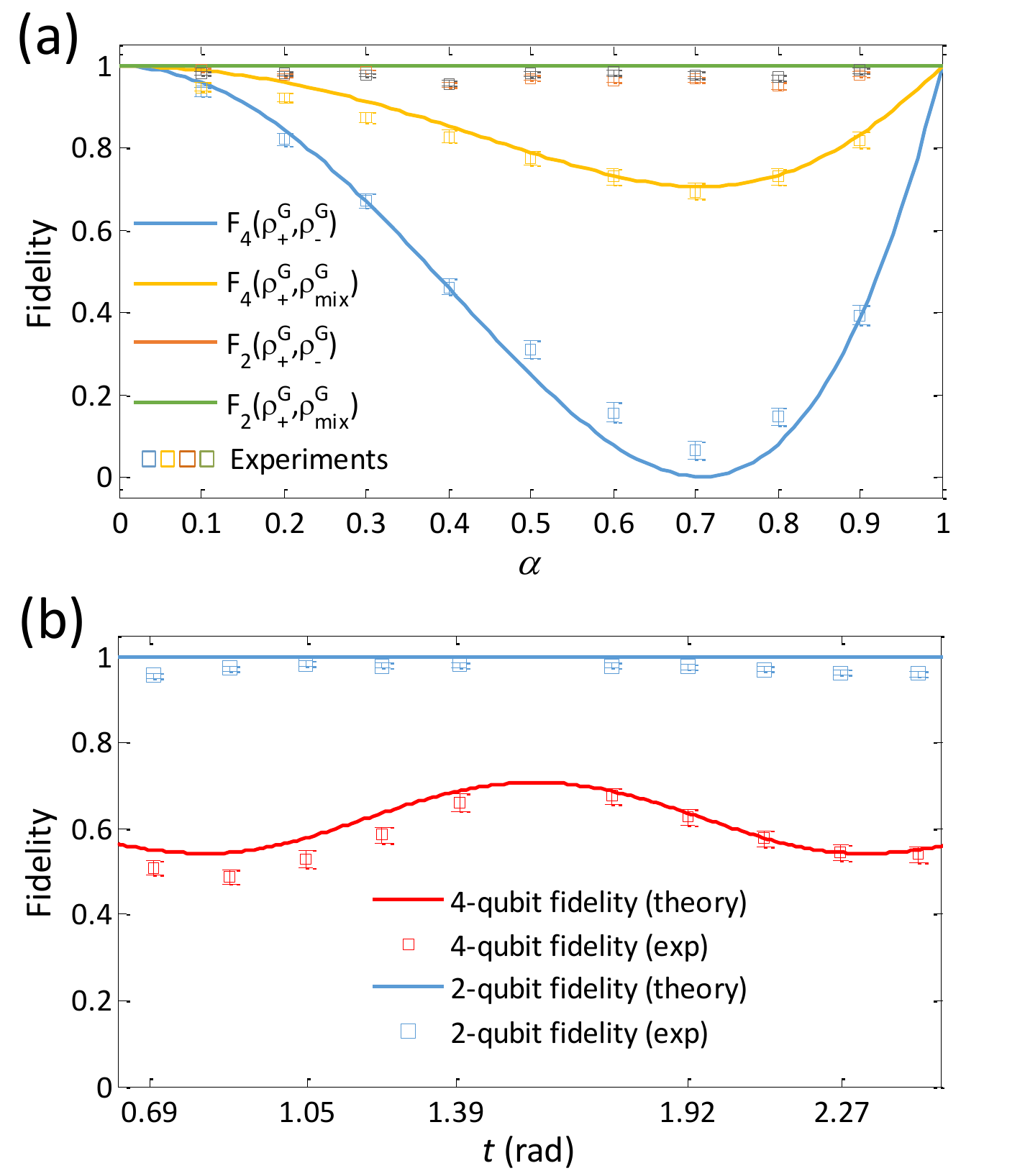}
\end{center}
\setlength{\abovecaptionskip}{-0.00cm}
\caption{\footnotesize{(a) GHZ-type states (Class A) such as $\rho^{\text{G}}_{+}$ in Eq. (\ref{ghz_exp}) are neither UDP nor UDA. The 4-qubit fidelities between  $\rho^{\text{G}}_{+}$ and $\rho^{\text{G}}_{-}$ (blue), and $\rho^{\text{G}}_{+}$ and $\rho^{\text{G}}_{mix}$ (yellow) are completely different, but they do have the same 2-RDMs (red and green, where the worst-case fidelity out of six possible 2-RDM fidelities is shown) up to minor experimental errors. The error bars are calculated from the imperfection of the GRAPE pulses and fitting procedure. (b) States in Class C are not UDA, so there can exist mixed states between which they have very low 4-qubit fidelity (red), but the same 2-RDMs (blue). However, these types of states are UDP so there do not exist any other 4-qubit pure states with the same 2-RDMs.}} \label{ghz1_results}
\end{figure}

Class A: States belonging to this class are neither UDP nor UDA by their 2-RDMs. The following states are in class A
\bea\label{ghz_exp}
& \ket{\text{GHZ}}_+  = \alpha\ket{0000}+\beta\ket{1111}, \\ \nonumber
& \ket{\text{GHZ}}_-   =  \alpha\ket{0000}-\beta\ket{1111}, \\ \nonumber
& \rho_{\text{mix}}^{\text{G}}  =  |\alpha|^2\ket{0000}\bra{0000}+|\beta|^2\ket{1111}\bra{1111},
\eea
and $\rho^{\text{G}}_{+}$ and $\rho^{\text{G}}_{-}$ are the density matrices of $\ket{\text{GHZ}}_+$ and $\ket{\text{GHZ}}_-$, respectively. All of these states have the same 2-RDMs, which means that the 2-RDMs are not sufficient to reconstruct these states. To verify this, we first need to prepare each state in Eq. (\ref{ghz_exp}) from $\ket{0000}$. For $\rho^{\text{G}}_{+}$, qubit 1 firstly undergoes a rotation around \emph{y}-axis that $R_y(\theta)=e^{-i \theta \sigma_y/2}$ with $\theta=2 \arccos(\alpha)$. Then three controlled-NOT (CNOT) gates CNOT$_{12}$, CNOT$_{13}$ and CNOT$_{14}$ are applied consecutively, where qubit 1 is the control and others are targets.  The single-qubit rotation $R_y(\theta)$ is realized by a 1 ms GRAPE pulse and the 3 CNOT gates are realized by a 30 ms GRAPE pulse. We can similarly construct $\rho^{\text{G}}_{-}$ by instead employing a single-qubit rotation of $R_y(-\theta)=e^{i \theta \sigma_y/2}$. For $\rho_{\text{mix}}^{\text{G}}$, we simply prepare a classical distribution of two pure states $\ket{0000}$ and $\ket{1111}$. In these experiments we prepare nine distinct states by varying $\alpha$ from 0.1 to 0.9 with 0.1 increment.

Class B: States belonging to this class are both UDP and UDA, with the W-type state in Eq. (\ref{wstate_e}) being a typical example. In experiment, we simply set $a=b$ and $c=d$, and then prepare six inputs $\ket{\text{W}}$ by changing $a$ from 0.1 to 0.6 with 0.1 increment.  This state preparation is directly realized by a state-to-state GRAPE pulse with a duration of 20 ms.

Class C: States belonging to this class are UDP but not UDA. The type of state we prepare, $\ket{\psi_S}$ is described in Eq. (\ref{target_state}) and conforms to the following parametrization
\begin{align*}\label{c0c2c4_t}
c_0&=\frac{\sin{t}-\sin{\theta}\cos{t}}{\sqrt{2}}, \\
c_2&=\cos{\theta}\cos{t}, \\
c_4&=-\frac{\sin{\theta}\cos{t}+\sin{t}}{\sqrt{2}},
\end{align*}
where we fix $\theta=\pi/12$ and choose $t$ from $\pi/6+\pi/18$ to $5\pi/6-\pi/18$, and increment by $\pi/18$. With the exception of the point $t=\pi/2$ this curve lies within the region of states that are  UDP but not UDA, as outlined in appendices A and B. All these states are prepared by state-to-state GRAPE pulses with a fixed duration of 20 ms. In order to demonstrate that these states are UDA we also prepare corresponding mixed states with the same 2-RDMs as outlined in appendix D \cite{supple}.

(iii) After preparing these states, we perform 4-qubit QST~\cite{leskowitz2004state,lee2002quantum}, which includes measuring the 2-RDMs. To determine the original 4-qubit state, a maximum likelihood approach \cite{altepeter2005photonic} is adopted to reconstruct the most likely state based on the measured 2-RDMs.

\textit{Results}---Now we discuss the effectiveness and stability of QST via 2-RDMs for each class of states.

\begin{figure}[htb]
\begin{center}
\includegraphics[width= 0.95\columnwidth]{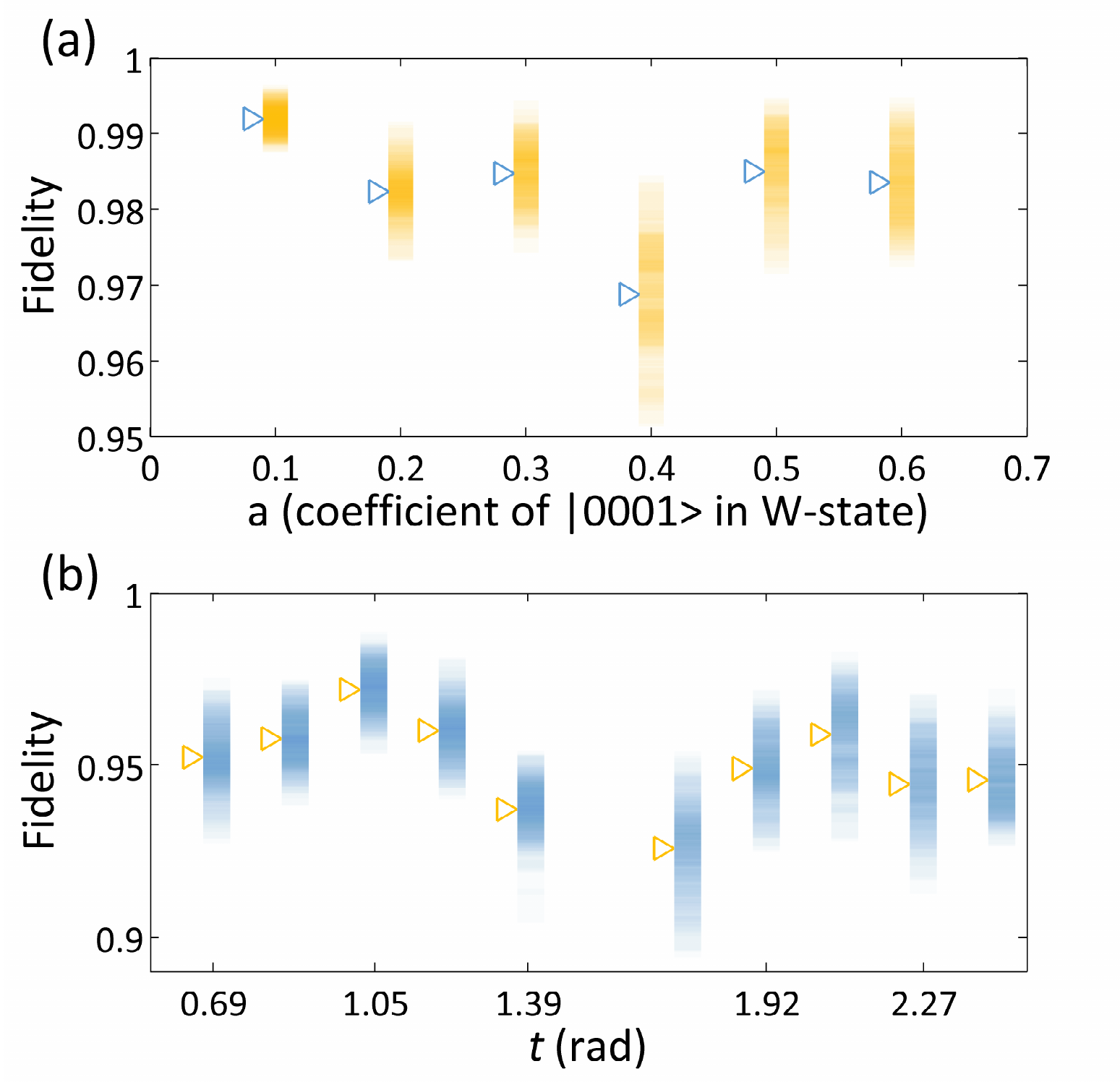}
\end{center}
\setlength{\abovecaptionskip}{-0.00cm}
\caption{\footnotesize{Stability test against experimental noise for $\ket{\text{W}}$ and $\ket{\psi_S}$. The noise is artificially added in Gaussian distribution to the measured 2-RDMs under experimental conditions, by randomly sampling 90 distinct sets of 2-RDMs. The arrows indicate the mean for each sampled results. (a) Fidelities of the $\ket{\text{W}}$ (Class B) in noisy environment. The $x$-axis is the coefficient $a$ defined in Eq. (\ref{wstate_e}). (b) Fidelities of the $\ket{\psi_S}$ (Class C) in noisy environment, as a function of $t$ defined in Eq. (\ref{target_state}).}} \label{stability}
\end{figure}

Class A: In Fig. \ref{ghz1_results}(a), it is clear that any two of $\rho^{\text{G}}_{+}$, $\rho^{\text{G}}_{-}$ and $\rho^{\text{G}}_{mix}$ have completely different fidelities in the 4-qubit form (blue and yellow), but they share the same 2-RDMs up to minor experimental errors (red and green). Therefore these states are neither UDP nor UDA, and it is insufficient to rely only on their 2-RDMs for QST.

Class B: The W-type state in Eq. (\ref{wstate_e}) is known to be UDA. In Fig. \ref{stability}(a), the blue triangles represent the fidelities $F(\rho_{\text{qst}}^{\text{W}},\rho_{\text{2rdm}}^{\text{W}})$ between the prepared 4-qubit state $\rho_{\text{qst}}^{\text{W}}$ via full QST and the reconstructed 4-qubit state $\rho_{\text{2rdm}}^{\text{W}}$ via 2-RDMs. For every tested W-type state, the worst fidelity is still about $97\%$ as shown by the triangles in Fig. \ref{stability}(a). This indicates that the 2-RDMs are indeed sufficient for the reconstruction of the original 4-qubit state.

However, under realistic experimental conditions, the prepared state $\rho_{\text{qst}}^{\text{W}}$ unavoidably deviates from the desired state. This may drive it outside the UDA region, so that it is no longer UDA. To test if this is the case, we simulate different outputs of 2-RDMs by adding Gaussian distributed noise and repeating the reconstruction of the 4-qubit state via the 2-RDMs, as outlined in appendix F \cite{supple}. From the yellow bars in Fig. \ref{stability}(a) it can be seen that even with artificial noise, QST via 2-RDMs is stable, since the fidelity is always over 0.95.

Class C: This class is UDP, which means we do not have any other pure state that gives the same 2-RDMs other than the target state. However, it is not UDA, so there do exist some mixed state (see appendix D \cite{supple}) with the same 2-RDMs. Fig. \ref{ghz1_results}(b) illustrates such results. Both in theory and experiment, we see that the target state $\ket{\psi_S}$ and a corresponding mixed state have low fidelity with one another (yellow), but the same 2-RDMs (blue). Therefore, when reconstructing this type of 4-qubit state via its 2-RDMs, we need to assume that the original state is pure. Otherwise it is likely to obtain some mixed state which will not necessarily be the true state of the system.

Similarly to the W-type state, we test whether the UDP property of $\ket{\psi_S}$ is stable against noise. As seen in Fig. \ref{stability}(b), even under the application of Gaussian noise, as long as we assume our state is pure we can always reconstruct the correct 4-qubit state with high fidelity (>0.90) using only its 2-RDMs.

\textit{Conclusion}---In summary, we disprove the hypothesis that UDP implies UDA for RDMs \cite{CJRZ12} by demonstrating the existence of a family of 4-qubit states that are UDP but not UDA by their 2-RDMs. This new finding allows us to classify pure states into three classes according to their UD properties, in order to improve the efficiency of QST: in Class A where the state is neither UDP nor UDA, full QST is necessary; in Class B where the state is UDP and UDA, the measurement of 2-RDMs is sufficient to determine the global state; in Class C where the state is UDP but not UDA, the measurement of 2-RDMs combined with the assumption that the global state is pure is sufficient. This approach simplifies QST significantly, since a full QST of $n$ qubits requires $4^n-1$ observables while 2-RDM measurement requires ${n \choose 1}\times 3+{n \choose 2}\times 9$ observables (all weight-1 and weight-2 Pauli operators) only.

We check the feasibility of this protocol for each class with a 4-qubit NMR quantum processor. The results indicate that for Classes B and C it is not necessary to implement the full QST---2-RDMs already enables the reproduction of the global state with high fidelities. As there are always experimental errors, we also demonstrate the stabilities of this protocol, namely, whether it is robust against experimental noise. The results reveal that the approach of doing QST solely via the measurement of 2-RDMs is robust to the noise under our experimental conditions, and hopefully behaves the same in other experimental platforms.

\begin{acknowledgments}
{\bf Acknowledgments.} We are grateful to the following funding sources: NSERC (D.L., N.Y., J.K., B.Z. and R.L.); Industry Canada (R.L.); CIFAR (B.Z. and R.L.); National Natural Science Foundation of China under Grants No. 11175094 and No. 91221205 (T.X. and G.L.); National Basic Research Program of China under Grant No. 2015CB921002 (T.X. and G.L.).
\end{acknowledgments}

\appendix
\section{Appendix A: Backgrounds of UDP vs UDA problem}
In this appendix, we go through a brief history of the UDP vs UDA problem.

If a pure state $\ket{\psi}$ is UDA then there do not exist any states, pure or mixed, having the same RDMs as $\ket{\psi}$, thus the set of states which are UDP ($S_{UDP}$) are a subset of the set of states which are UDA ($S_{UDA}$). We can also define $S_{UDk}$ to be the set of states which are uniquely determined among states with rank no more than $k$. We therefore have the hierarchy $S_{UDA}=S_{UDn} \subseteq \cdots \subseteq S_{UDk} \subseteq \cdots \subseteq S_{UD1}=S_{UDP}$ where $n$ is the dimension of the whole system. It is not clear when this hierarchy collapses. If any of these inclusions turned out to be strict it would follow that $S_{UDA}\neq S_{UDP}$, which would intuitively seem to be the case.

However, in \cite{Linden2002}, it was demonstrated that all three qubit pure states are UDA except for the GHZ type state $a\ket{000}+b\ket{111}$. Clearly, any states of the form $ae^{i\theta_1}\ket{000}+be^{i\theta_2}\ket{111}$ have the same $2$-RDMs, and so are not UDP. In other words, for three qubit system, if a state is not UDA by its $2$-RDMs, then neither is it UDP. This observation leads us to the conclusion that UDP implies UDA in three-qubit system.

Furthermore, in \cite{Linden2002b, JL05}, it was proved that generic $N$-party pure quantum states are UDA by the RDMs of just over half the parties. Unfortunately it is difficult to characterize all the non-UDA states, save for some well-known exceptions. For example, the $N$-party GHZ states $\ket{0}^{\otimes N}+e^{i\theta}\ket{1}^{\otimes N}$, which are not UDP.

All of these facts suggest that perhaps UDP does imply UDA for general multipartite systems. For example Ref. \cite{Chen2013} gives an extensively study of the relationship between UDA and UDP in a more general setting, where RDMs are replaced by general observables. There it has been proved that under some restriction of the observables, UDP implies UDA.

In \cite{BDK15,KB15,Chen2013}, the relations between $S_{UDk}$ and $S_{UDA}$ are studied, but with different notations and in a different setting. Instead of RDMs, general POVM measurements are allowed, and they are mainly focus on the structures of measurements by which all states with rank $\leq k$ are $S_{UDk}$ or $S_{UDA}$, respectively.

\section{Appendix B:Proof of UDP}
In this appendix, we deal with steps (1) and (2) by constructing a class of four-qubit UDP symmetric pure states.

First we show that almost all 4 qubit pure symmetric states of the form:
\begin{equation}\label{c0c2c4}
\ket{\psi_S}=c_0 \ket{w_0}+c_2 \ket{w_2} +c_4\ket{w_4} \\
\end{equation}
for $c_i \in \mathbb{R}$ are UDP by their $2$-RDMs, where $\ket{w_j}$ is the normalized state that is proportional to $P_{\textrm{sym}}\bigl( \ket{0}^{\otimes j} \otimes \ket{1}^{\otimes 4-j} \bigr)$ with $P_{\textrm{sym}}$ being the projection onto the 4-qubit symmetric subspace.

Consider the Schmidt decomposition of $\ket{\psi_S}$ between the 1,2 and 3,4 Hilbert spaces. Note that it is symmetric under the exchange of the 1,2 and 3,4 Hilbert spaces, it is straightforward to see that if there is no degeneracy in the Schmidt coefficients then the Schmidt decomposition must take this form:
\begin{equation}\label{psi}
\ket{\psi_S}=\sum_{i=1}^3 \sqrt{\lambda_i} \ket{\mu_i}\ket{\mu_i},
\end{equation}
where $\lambda_i\neq \lambda_j$ for $i\neq j$.

We assume there is another symmetric pure state $\ket{\phi_S}=\sum b_i\ket{w_i}$ which has the same $2$-RDMS of $\ket{\psi_S}$.
That is, $$\tr_2[\psi_S]=\tr_2[\phi_S]=\sum_{i=1}^3 \lambda_i \ketbra{\mu_i}{\mu_i},$$ with $\tr_2$ denoting operator tracing out two qubit subsystems. One can observe that for $\ket{\phi_S}$ to be distinct there must exist non-trivial phases $e^{i\theta_j}$ such that
\begin{equation}\label{phi}
\ket{\phi_S}=\sum_{j=1}^3 \sqrt{\lambda_j} e^{i\theta_j} \ket{\mu_j}\ket{\mu_j}.
\end{equation}
Note that $\ket{\mu_i}$ must be symmetric under the exchange of particles 1 and 2, and so can be expanded into a symmetric basis:
\begin{equation} \label{mu}
\ket{\mu_i}=\sum_{k=0}^2 \alpha_k^i \ket{s_k}
\end{equation}
where
\begin{align*}
\ket{s_0}&=\ket{00} \\
\ket{s_1}&=\frac{1}{\sqrt{2}}(\ket{10}+\ket{01})  \\
\ket{s_2}&=\ket{11}.
\end{align*}
Expanding equations \ref{psi} and \ref{phi} in terms of equation \ref{mu} gives the following expressions:
\begin{align*} \label{phiExpansion}
\ket{\psi_S}=&\sum_{i=0}^2\sum_{j\geq i}^2 A_{ij} \ket{s_{ij}}, \\
\ket{\phi_S}=&\sum_{i=0}^2\sum_{j\geq i}^2 B_{ij} \ket{s_{ij}}, \\
\ket{s_{ij}}=&\ket{s_i}\ket{s_j}+(1-\delta_{i j})\ket{s_j}\ket{s_i},\\
A_{ij}=\sum_{k=1}^3  \sqrt{\lambda_k} &\alpha_i^k  \alpha_{j}^k \text{ , }B_{ij}=\sum_{k=1}^3  \sqrt{\lambda_k} e^{i\theta_k}\alpha_i^k  \alpha_{j}^k,
\end{align*}
where $\delta_{i j}$ is Kronecker symbol.
Any symmetric state $\ket{\psi_S}$ must satisfy the following equality:
\begin{equation} \label{condition}
S_{23} \ket{\psi_S}- \ket{\psi_S}=0,
\end{equation}
where $S_{23}$ denotes the swap gate on the second qubit and the third qubit.

One can verify the following
\begin{align*}
S_{23}\ket{s_{00}}&=\ket{s_{00}}\\
S_{23}\ket{s_{01}}&=\ket{s_{01}}\\
S_{23}\ket{s_{02}}&=\ket{s_{11}}+\ket{as_{11}}\\
S_{23}\ket{s_{11}}&=\frac{1}{2}(\ket{s_{02}}+\ket{s_{11}}-\ket{as_{11}})\\
S_{23}\ket{s_{12}}&=\ket{s_{12}}\\
S_{23}\ket{s_{22}}&=\ket{s_{22}}\\
\ket{as_{11}}&:=\left(\frac{\ket{0}\ket{1}-\ket{1}\ket{0}}{\sqrt{2}}\right) \left(\frac{\ket{0}\ket{1}-\ket{1}\ket{0}}{\sqrt{2}}\right).
\end{align*}
Applying $S_{23}$ to $\ket{\psi_S}$:
\begin{align*}
&\ket{\psi_S}-S_{23}\ket{\psi_S}=\left(A_{02}-\frac{A_{11}}{2}\right)\ket{s_{02}}+\\
&+\left(\frac{A_{11}}{2} -A_{02} \right)\ket{s_{11}}+\left(\frac{A_{11}}{2}-A_{02}\right)\ket{as_{11}}\\
\end{align*}
It is clear $\ket{\psi_S}$ and $\ket{\phi_S}$ satisfy the above condition if and only if $A_{02}=\frac{A_{11}}{2}$ and $B_{02}=\frac{B_{11}}{2}$

By examining the Schmidt decomposition of $\ket{\psi_S}$, we note that $\ket{s_1}$ is an eigenvector of the 2-RDM of $\ket{\psi_S}$. Without loss of generality, we assume $\ket{\mu_1}=\ket{s_1}$. This implies $A_{11}=\sqrt{\lambda_1}$, and up to a global phase $B_{11}=\sqrt{\lambda_1}$. Therefore $A_{02}=B_{02}$, or in expanded form:
\begin{equation} \label{cond2}
 \sqrt{\lambda_2} \alpha_0^2 \alpha_2^2 +\sqrt{\lambda_3} \alpha_0^3 \alpha_2^3= \sqrt{\lambda_2} e^{i \theta_2}\alpha_0^2 \alpha_2^2 +\sqrt{\lambda_3} e^{i \theta_3} \alpha_0^3 \alpha_2^3.
 \end{equation}
Let $V_a=\sqrt{\lambda_2} \alpha_0^2 \alpha_2^2$ and $V_b=\sqrt{\lambda_3} \alpha_0^3 \alpha_2^3$ be vectors on the 2 dimensional complex plane. Consider two circles A and B. Let circle A be centered at the origin with radius $|V_a|$ and let circle B be centered at $V_a+V_b$ with radius $|V_b|$. Clearly circle A and circle B intersect at the point $V_a$. Asking how many solutions there are to equation \ref{cond2} corresponds to asking how many times circle A and circle B intersect.

\begin{figure}[h]
\includegraphics[scale=.6]{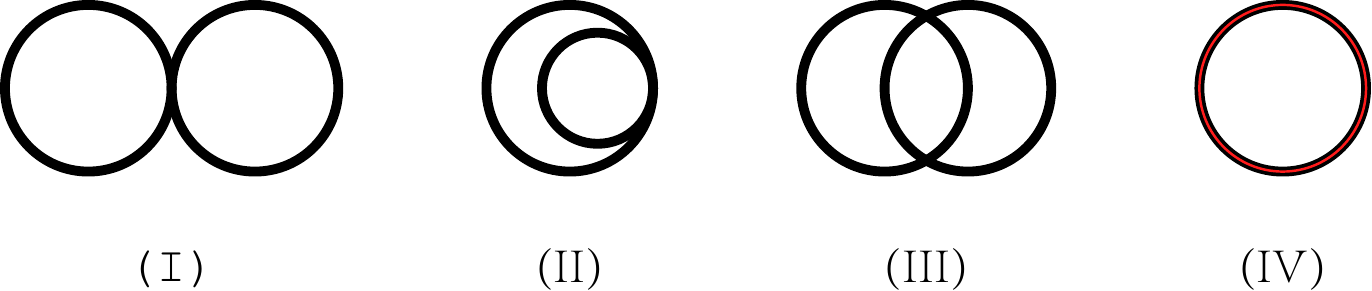}
\caption{Classes of circle intersections}
\label{circles}
\end{figure}

There are 4 distinct arrangements of circles, as illustrated in figure \ref{circles}. Under condition (I) $arg(V_a)=arg(V_b)$ and there is a single intersection; (II) $arg(V_a)=arg(V_b)+\pi$, $|V_a| \neq |V_b|$ and there is one intersection; (III)  $arg(V_a)-arg(V_b) \neq 0,\pi$ and there are two intersections; (IV) $arg(V_a)=arg(V_b)+\pi$, $|V_a| = |V_b|$ and there are an infinite number of intersections.

Equation \ref{cond2} has only one solution, namely $e^{i\theta_2}=e^{i\theta_3}=1$ if and only if either case (I) or case (II) is true.

Since finding the Schmidt decomposition for $\ket{\psi_S}$ is equivalent to solving the eigensystem of a real symmetric matrix, as long as $\lambda_2 \neq \lambda_3$ we are free to let $\alpha_i^j$ and $\sqrt{\lambda_j}$ be real without loss of generality. This rules out case (III).

According to $\ket{\mu_1}=\ket{s_1}$, we know that:
\begin{align*}
\ket{\mu_2}=\alpha_0^2 \ket{s_0} +\alpha_2^2 \ket{s_2}\\
\ket{\mu_3}=\alpha_0^3 \ket{s_0} +\alpha_2^3 \ket{s_2}.
\end{align*}
Using the fact that these two vectors are real and orthonormal, it is straightforward to show that $|\alpha_0^3|=|\alpha_2^2|$ and $|\alpha_2^3|=|\alpha_0^2|$, which implies that if $|\lambda_2| \neq |\lambda_3|$ then case (IV) is ruled out.

Therefore as long as $\lambda_2 \neq \lambda_3$ equation \ref{cond2} has a single solution: $e^{i\theta_2}=e^{i\theta_3}=1$, then $\ket{\phi_S}=\ket{\psi_S}$. That is, $\ket{\psi_S}$ is uniquely determined by its 2-RDMS among all symmetric pure states $\ket{\phi_S}$.

In the following, we show the validity of step (2), that is, any pure state which is uniquely determined
among all symmetric pure states is also uniquely determined
among all pure states (UDP).

Suppose any two local states $\rho_{i,j}$ of multipartite state $\rho_{1,2,\cdots, n}$ lives in the bipartite symmetric space. Then we can conclude that $\rho_{1,2,\cdots n}$ must live in the $n$ partite symmetric space. To see this, we notice that $Q_{i,j}\rho_{i,j}=0$ for all $1\leq i<j\leq r$ with $Q_{i,j}$ being the projection onto the antisymmetric subspace of the Hilbert space of particles $i,j$. Then, we have $Q_{i,j}\rho_{1,2,\cdots, n}=0$ for all $1\leq i<j\leq n$. Therefore, $S_{i,j}\rho_{1,2,\cdots, n}=\rho_{1,2,\cdots, n}$ for all $1\leq i<j\leq n$ with $S_{i,j}$ being the SWAP operator of the particles $i,j$.

That completes the proof that all pure symmetric states of the form \ref{c0c2c4} are UDP save for those states whose 2-RDMs are degenerate.

\section{Appendix C: Proof of not UDA}
Here we perform step (3), determining the regime where $\ket{\psi_S}$ is not UDA.
We observe that if the 2-RDM $\rho_2$ of $\ket{\psi_S}$ is separable, then it can be expressed as $\rho_2=\sum_i p_i \alpha_i^{\otimes 2}$. Then we can construct a state $\rho_4=\sum_i p_i \alpha_i^{\otimes 4}$ which has the same 2 particle reduced density matrices as $\ket{\psi_S}$. Therefore, if $\ket{\psi_S}$ has a separable 2-RDM whose rank is not 1, then it is not uniquely determined among all quantum states by its 2-RDMS(2-UDA). A 2 qubit state is separable if it has a positive partial transpose (PPT)\cite{Horodecki1996}. Direct calculation shows that a 2-RDM of $\ket{\psi}$ is PPT when
\begin{align}\label{ppt}
c_2^4/9 &\leq (c_0^2+c_2^2/6)(c_4^2+c_2^2/6) \\
c_2/3 &\geq c_0 /\sqrt{6} + c_4 / \sqrt{6} \nonumber
\end{align}

The results of appendices A and B are summarized in Fig.~\ref{sphere} where we have illustrated a regime where the state $\ket{\psi_S}$ is uniquely determined among all pure states by its 2-RDMS(2-UDP) but not 2-UDA. Given that we are considering a unit length vector of three real parameters $(c_0$, $c_2$,$c_4)$, we can map our parameter space to the surface of a sphere. The green region illustrates the domain where the 2-RDMs of $\ket{\psi_S}$ are separable, and thus $\ket{\psi_S}$ is not 2-UDA. The red curves indicate where $\ket{\psi_S}$ fails to be 2-UDP. Thus the green domain not intersecting the red curves is where $\ket{\psi_S}$ is 2-UDP but not 2-UDA. In experimentally reproducing these states, we should like to have a one-parameter family of curves well within the 2-UDP but not 2-UDA states. The family we consider in this paper is illustrated by the blue curve.

\begin{figure}[h]\label{sphere}
\includegraphics[scale=.6]{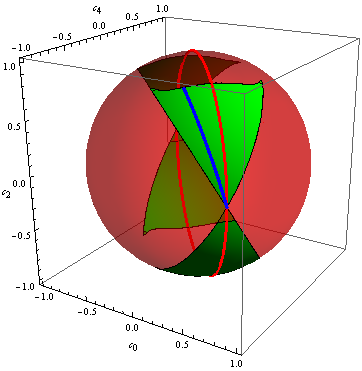}
\caption{The parameter space of $\ket{\psi}$. The green region is where $\ket{\psi_S}$ is 2-UDP and not 2-UDA, given by the inequalities \ref{ppt}. The red region is where $\ket{\psi}$ is not 2-UDP and/or may be 2-UDA. The blue line includes the set of experimentally prepared states.}
\label{sphere}
\end{figure}

Information-theoretically, if we do not assume that the state compatible with the $2$-RDMs is pure, then the best possible inference of the state is the one of maximum entropy compatible with the $2$-RDMs. Such a state is unique. To compute what such a state theoretically should be we employ a variational algorithm. The state of maximum entropy compatible with the 2-RDMs $\rho^*$ should live on the set of thermal states given by $\rho(H)=\frac{e^{-H}}{Tr[e^{-H}]}$, where $H$ is a Hermitian trace zero operator which acts only 2-locally. This set can be searched by starting with a guess state given by the maximally mixed state, and then iteratively updating our guess by minimizing the Hilbert-Schmidt distance from some state $(1-\alpha) \mathds{1}/d + \alpha \ketbra{\psi_S}{\psi_S}$ where at each iteration $\alpha$ approaches $1$.

\section{Appendix D: The separable decomposition}
In this appendix, we construct a separable decomposition of the 2-RDMs $\rho_2=\sum_i p_i \alpha_i^{\otimes 2}$ of $\ket{\psi_S}$. With a separable decomposition we can then construct a mixed state $\rho_4=\sum_i p_i \alpha_i^{\otimes 4}$ with the same 2-RDMs as $\ket{\psi_S}$. Employing Lorentz transformation techniques similar to those outlined here \cite{Leinaas2006} we can construct the following separable representation for any of the states satisfying condition \ref{ppt}.

\begin{equation}
\rho_2= \sum_{k\in \{x,y,z\}} \frac{\left(p_k^+ \alpha_k^+ \otimes \alpha_k^+ + p_k^- \alpha_k^- \otimes \alpha_k^- \right)}{2}.
\end{equation}
The derivation of this construction is as follows.
Consider the 2-RDM of our state $\ket{\psi_S}$:
\begin{align*}\rho_2=\left(
\begin{array}{cccc}
 c_0^2+\frac{c_2^2}{6} & 0 & 0 & \frac{c_0 c_2}{\sqrt{6}}+\frac{c_4 c_2}{\sqrt{6}} \\
 0 & \frac{c_2^2}{3} & \frac{c_2^2}{3} & 0 \\
 0 & \frac{c_2^2}{3} & \frac{c_2^2}{3} & 0 \\
 \frac{c_0 c_2}{\sqrt{6}}+\frac{c_4 c_2}{\sqrt{6}} & 0 & 0 & \frac{c_2^2}{6}+c_4^2 \\
\end{array}
\right)\end{align*}
To simplify notation we can define the variables $a,b,c,d$ such that:
\begin{align*}\rho_2=\left(
\begin{array}{cccc}
 a & 0 & 0 & c \\
 0 & d & d & 0 \\
 0 & d & d & 0 \\
 c & 0 & 0 & b \\
\end{array}
\right).\end{align*}
$\rho_2$ can then be expressed in its Pauli operator expansion as:
\begin{align*}
\rho_2=&\frac{1}{4}\mathds{1}\otimes \mathds{1}+ \frac{c+d}{2} \sigma_x \otimes \sigma_x +\frac{d-c}{2} \sigma_y \otimes \sigma_y \\
&+ \frac{a+b-2d}{4} \sigma_z \otimes \sigma_z + \frac{a-b}{4} \left(\sigma_z \otimes \mathds{1} + \mathds{1} \otimes \sigma_z \right)
\end{align*}
Consider now the following transformation:
\begin{align*}
\tilde{\rho_2}=&\left(V \otimes V \right) \rho_2 \left(V \otimes V \right)\\
V =& \left(
\begin{array}{cc}
 e^{\epsilon} & 0 \\
 0 & e^{-\epsilon} \\
\end{array}
\right)
\end{align*}

We can again write the Pauli expansion:

\begin{align*}
\tilde{\rho_2}&=\frac{a e^{4\epsilon}+2d+b e^{-4\epsilon}}{4}\mathds{1}\otimes \mathds{1}+ \frac{c+d}{2} \sigma_x \otimes \sigma_x +\frac{d-c}{2} \sigma_y \otimes \sigma_y \\
&+ \frac{ae^{4\epsilon}-2d+b e^{-4\epsilon}}{4} \sigma_z \otimes \sigma_z + \frac{ae^{4\epsilon}-b e^{-4\epsilon}}{4} \left(\sigma_z \otimes \mathds{1} + \mathds{1} \otimes \sigma_z \right)
\end{align*}

We are free to choose $\epsilon$ such that $e^{4\epsilon}=\sqrt{\frac{b}{a}}$. Therefore,
\begin{align*}
\tilde{\rho_2}=\frac{1}{4}\left(d_0 \mathds{1}\otimes \mathds{1}+ d_x \sigma_x \otimes \sigma_x +d_y \sigma_y \otimes \sigma_y + d_z \sigma_z \otimes \sigma_z, \right)\end{align*}
where
\begin{align*}
d_0&=2\sqrt{ab}+2d \text{, } d_x=2c+2d,\\
d_y&=2d-2c \text{, }d_z=2\sqrt{ab}-2d,
\end{align*}
We can now note the following equality:
\begin{align*}
\tilde{\rho_2}=\sum_{k \in (x,y,z)} \frac{d_k}{8} \left( \left(\mathds{1}+\sigma _k\right)\otimes \left(\mathds{1}+\sigma _k\right)+\left(\mathds{1}-\sigma _k\right)\otimes \left(\mathds{1}-\sigma _k\right)\right)
\end{align*}

We can now perform the inverse map to retrieve our original state, noticing that our map preserves the manifestly separable structure of our state:
$$\rho_2=\left(V^{-1}\otimes V^{-1}\right)\tilde{\rho} \left(V^{-1}\otimes V^{-1}\right).$$

\section{Apendix E: Experimental fidelities for prepared states}

Experimentally, we use a 4-qubit sample $^{13}$C-labeled trans-crotonic acid dissolved in d6-acetone. In this appendix, we exhibit the fidelities of all prepared states in the 4- and 2-qubit manners.

\begin{figure}[htb!]
\begin{center}
\includegraphics[width= 1.1\columnwidth]{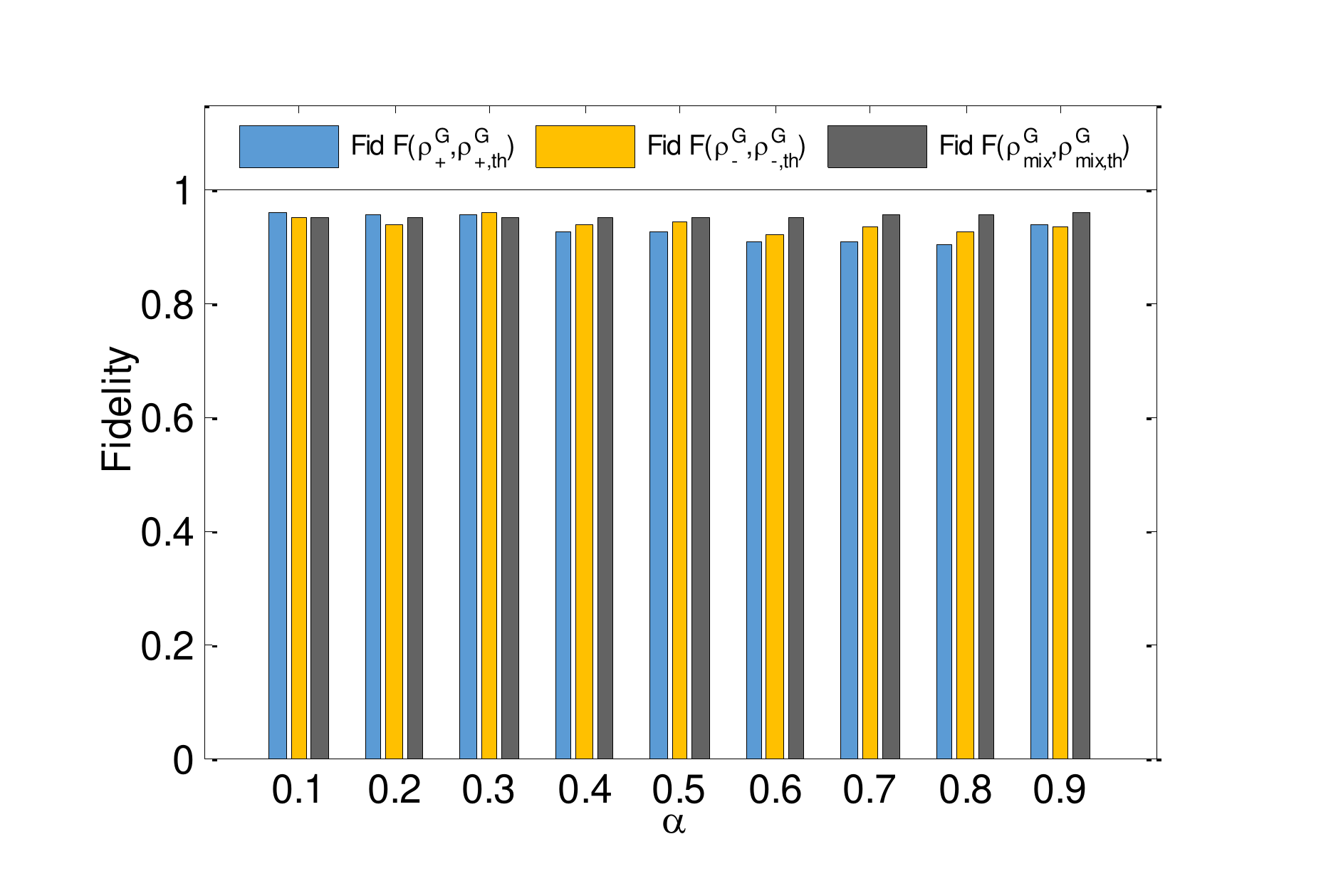}
\end{center}
\setlength{\abovecaptionskip}{-0.00cm}
\caption{\footnotesize{Fidelities for $\rho_{+}^{\text{G}}$, $\rho_{-}^{\text{G}}$ and $\rho_{\text{mix}}^{\text{G}}$, where $\rho_{+}^{\text{G}} = \ket{\text{GHZ}}_+\bra{\text{GHZ}}_+$ and $\rho_{-}^{\text{G}} = \ket{\text{GHZ}}_-\bra{\text{GHZ}}_-$. We prepare nine input states for each state by varying  $\alpha$ from 0.1 to 0.9 with 0.1 increment. The bars show the fidelities between them and the corresponding theoretical states, as a function of $\alpha$. The subscript $th$ means the corresponding theoretical state.}}\label{GHZexpFid}
\end{figure}

Firstly, we experimentally prepare the following states
\bea
& \ket{\text{GHZ}}_+  = \alpha\ket{0000}+\beta\ket{1111}, \\ \nonumber
& \ket{\text{GHZ}}_-   =  \alpha\ket{0000}-\beta\ket{1111}, \\ \nonumber
& \rho_{\text{mix}}^{\text{G}}  =  |\alpha|^2\ket{0000}\bra{0000}+|\beta|^2\ket{1111}\bra{1111},
\eea
where $\ket{\text{GHZ}}_+$ is neither UDP nor UDA, since there exists a pure state $\ket{\text{GHZ}}_-$ and a mixed state $\rho_{\text{mix}}^{\text{G}}$ which have the same 2-RDMs. $\ket{\text{GHZ}}_+$ and $\ket{\text{GHZ}}_-$ are prepared by GRAPE pulses after a the PPS preparation.  For $\rho_{\text{mix}}^{\text{G}}$, we create two components $\ket{0000}$ and $\ket{1111}$ respectively, and add them classically. The 4-qubit fidelities between all prepared states and the theoretical states are illustrated in Fig. \ref{GHZexpFid}, with all fidelities defined by $F(\rho,\sigma)=|Tr(\rho\sigma)|/\sqrt{Tr(\rho^2)}\sqrt{Tr(\sigma^2)}$.

\begin{figure}[htb!]
\begin{center}
\includegraphics[width= 0.85\columnwidth]{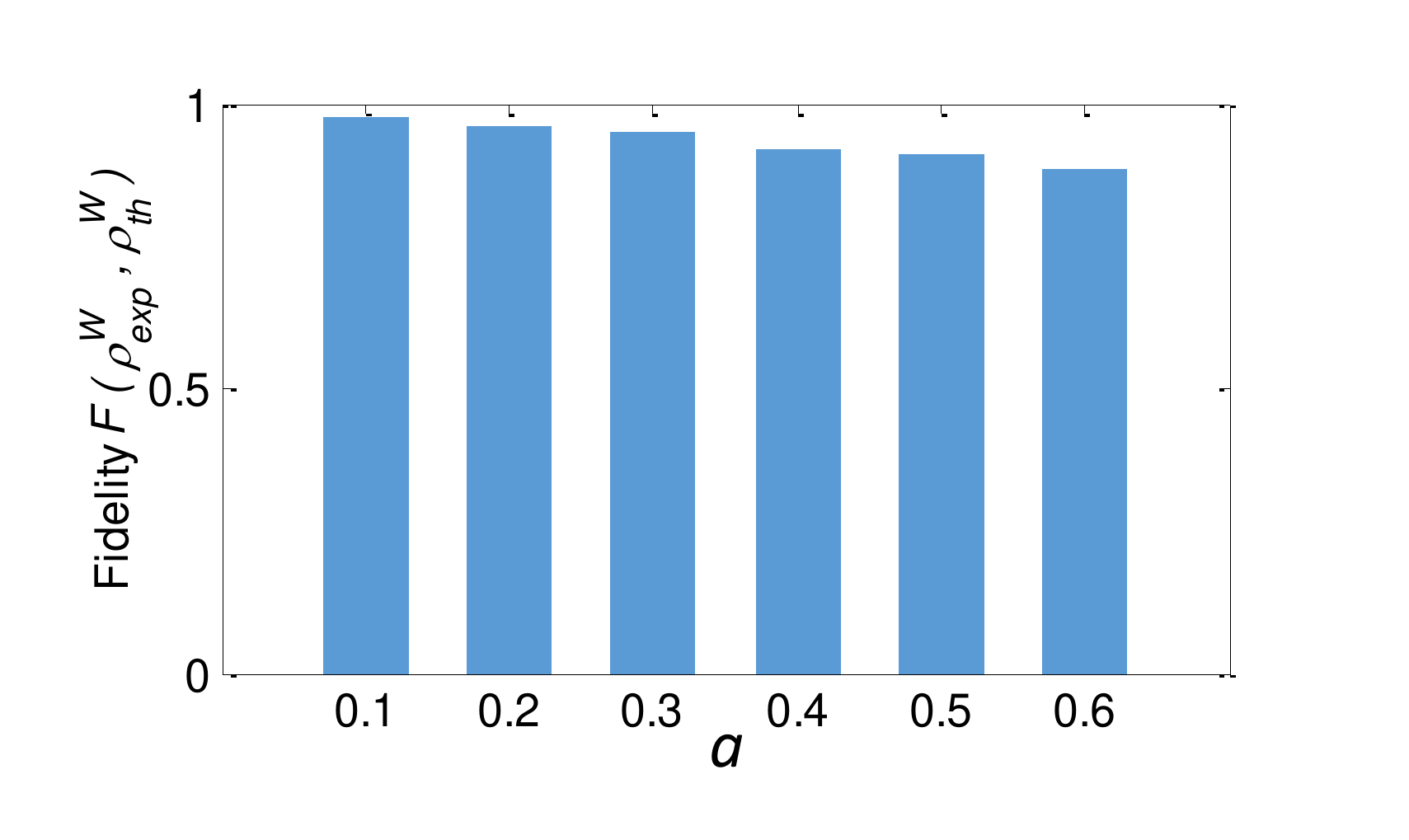}
\end{center}
\setlength{\abovecaptionskip}{-0.00cm}
\caption{\footnotesize{Fidelities between prepared $\rho^\text{W}_{exp}$ and the theoretical ones.  The bars show the fidelities between them and the corresponding theoretical states, as a function of $a$.}}\label{WexpFid}
\end{figure}

Secondly, we prepare the following states,
\begin{equation}
\ket{\text{W}}=a\ket{0001}+b\ket{0010}+c\ket{0100}+d\ket{1000},
\end{equation}
where we choose $a=b$ and $c=d$, and vary $a$ from 0.1 to 0.6 with 0.1 increment.  Fig. \ref{WexpFid} shows six 4-qubit fidelities between all prepared states $\ket{\text{W}}$ and the theoretical ones.

\begin{figure}[htb!]
\begin{center}
\includegraphics[width= 1.1\columnwidth]{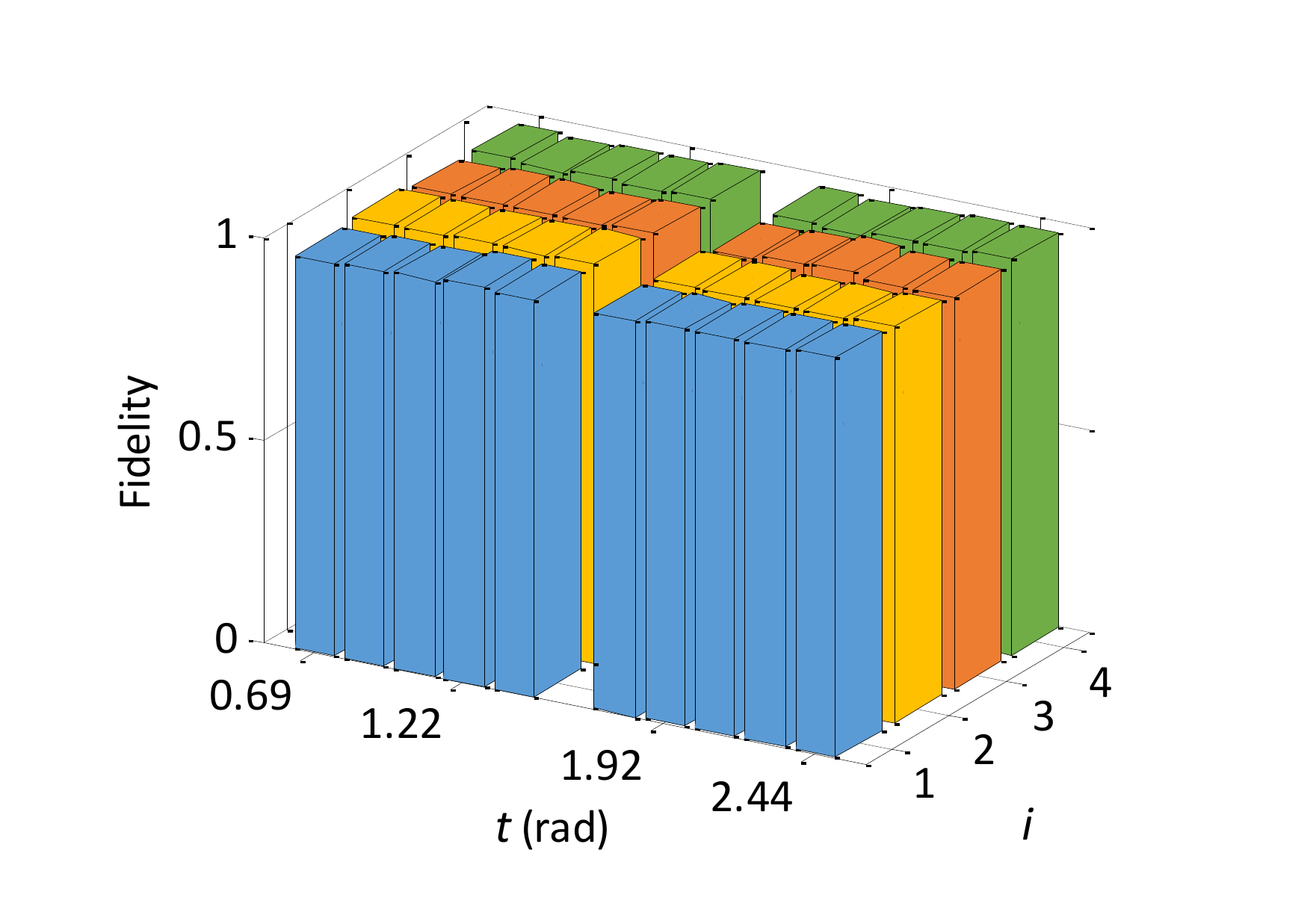}
\end{center}
\setlength{\abovecaptionskip}{-0.00cm}
\caption{\footnotesize{Fidelities for each prepared $\alpha_i \alpha_i \alpha_i \alpha_i$ for every $t$ in Eq. (\ref{c0c2c4_t}). The bars show the fidelities of each prepared $\alpha_i \alpha_i \alpha_i \alpha_i$ as a function of $t$ and $i$.}}\label{single_f}
\end{figure}

\begin{figure}[htb!]
\begin{center}
\includegraphics[width= 1.1\columnwidth]{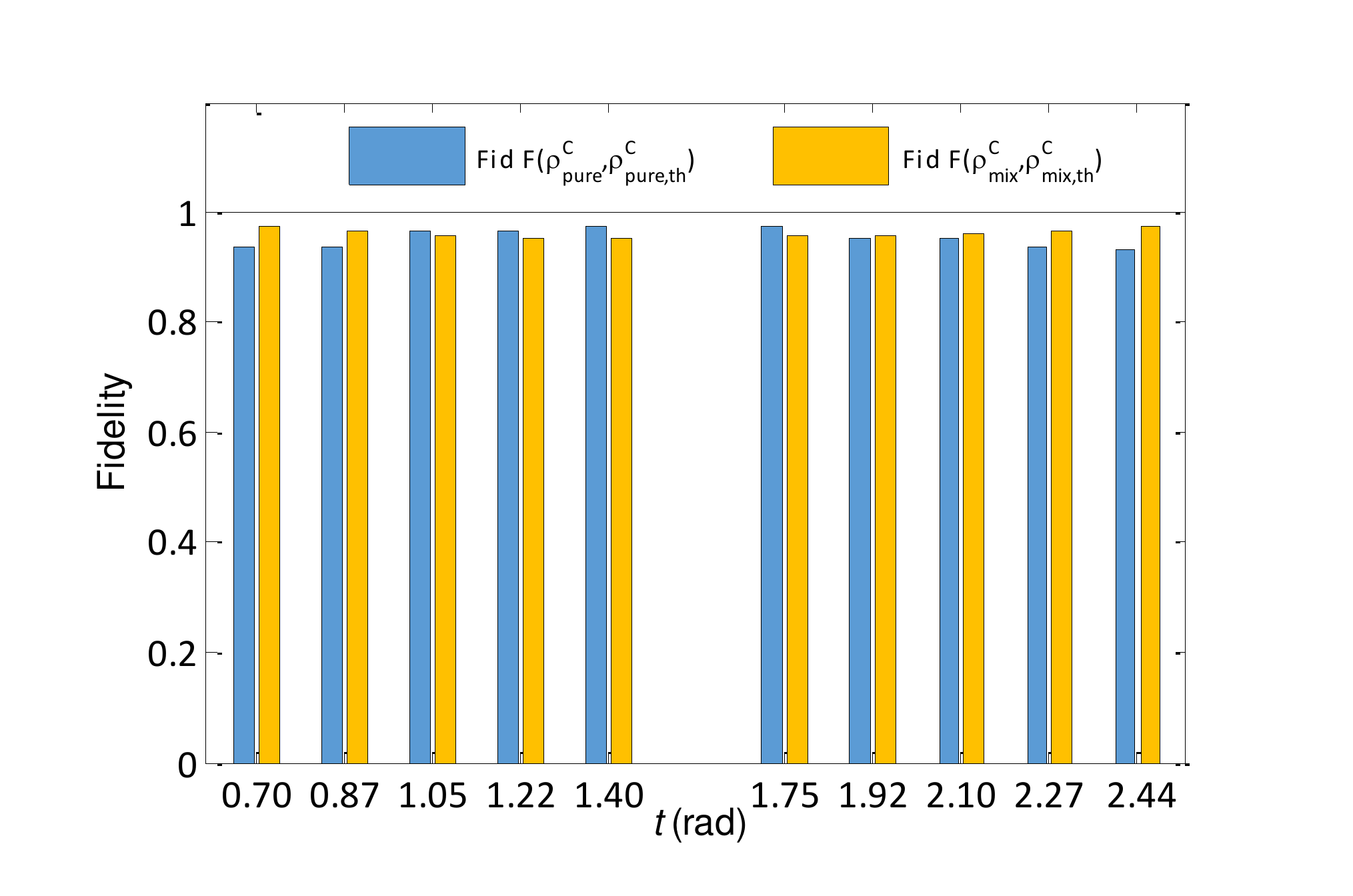}
\end{center}
\setlength{\abovecaptionskip}{-0.00cm}
\caption{Fidelities for prepared $\rho^C_{pure}$ and $\rho^G_{mix}$. In experiments, ten $\rho^{C}_{pure}$ and $\rho^{C}_{mix}$ are created by varying  $t$. }\label{c2expFid}
\end{figure}

Finally, we prepare the UDP but not UDA states $\ket{\psi_S}$
\begin{align*}
\ket{\psi_S}=c_0 \ket{w_0}+c_2 \ket{w_2} +c_4\ket{w_4} , c_0,c_2,c_4 \in \mathbb{R}
\end{align*}
where
\begin{align}\label{c0c2c4_t}
c_0=\frac{\sin{t}-\sin{\theta}\cos{t}}{\sqrt{2}}\\ \nonumber
c_4= -\frac{\sin{\theta}\cos{t}+\sin{t}}{\sqrt{2}}\\
c_2=\cos{\theta}\cos{t}. \nonumber
\end{align}
We fix $\theta=\pi/12$ and choose $t$ from $\pi/6+\pi/18$ to $5\pi/6-\pi/18$ with step size $\pi/18$ except the $t=\pi/2$ point. Experimentally, total ten input states $\rho^C_{pure} = \ket{\psi_S}\bra{\psi_S}$ were created. Meanwhile, we prepared a mixed state $\rho^C_{mix}$ which has the same 2-RDMs. $\rho^C_{mix}=\sum_i p_i \alpha_i \alpha_i \alpha_i \alpha_i$, where $\alpha_i$ is single-qubit density matrix and $p_i$ is the corresponding amplitude (details in Appendix C). For each $\rho^C_{mix}$, four more separable states $\alpha_i \alpha_i \alpha_i \alpha_i$ ($i=1,2,3,4$) excluding the $\ket{0000}$ and \ket{1111} are necessary to be created. Hence, in total 40 separable states are prepared experimentally. The fidelities of these 40 separable states are illustrated in Fig. \ref{single_f}.  In experiments, we respectively create each component $\alpha_i \alpha_i \alpha_i \alpha_i$, and then summarize over all components according to the coefficient $p_i$ to realize the mixed state $\rho^C_{mix}$. Fig. \ref{c2expFid} shows the fidelities for prepared $\rho^C_{pure}$ and $\rho^G_{mix}$.

\section{Appendix F: Add of Gaussian distributed noise}
In our experiment, the actual results deviate from the desired states due to some errors, such as the imperfections of GRAPE pulse, and the decoherence effect. In order to estimate the influence of the potential errors on existing experimental results, it is necessary to mimic the noise artificially according to the experimental errors. Now we introduce the method of adding Gaussian distributed noise in this experiment. For a 4-qubit state $\rho$, it can be expanded by Pauli basis.
\begin{equation}
\rho=\frac{\mathbb{I}}{16}+\sum_{i=1}^{255} M_i B_i,  B_i=\sigma_1\sigma_2\sigma_3\sigma_4,
\end{equation}
where $\sigma_1,\sigma_2,\sigma_3$ and $\sigma_4 \in {I, X,Y,Z}$, and $M_i$ is the expectation value of $B_i$.  In experiment, $\rho$ changes to $\rho'$ due to the errors:
\begin{equation}
\rho'=\frac{\mathbb{I}}{16}+\sum_{i=1}^{255} (M_i+e_i) B_i
\end{equation}
where $e_i$ is some error value originated from the experimental noise. In this experiment, the error model of $e_i$ can be described as
\begin{equation}
P(e_i)=\frac{3d}{c\sqrt{2\pi}} e^{-\frac{(3d)^2e_i^2}{2c^2}},
\end{equation}
where $P_i$ is a Gaussian distribution of $e_i$, and $d=16$ for a 4-qubit system.  The error percentage $c$ equals to about $11\%$ based on the estimation of our experimental noise. This model represents a random Gaussian distribution with zero mean and $99\%$ confidence.


\begin{thebibliography}{48}%
\makeatletter
\providecommand \@ifxundefined [1]{%
 \@ifx{#1\undefined}
}%
\providecommand \@ifnum [1]{%
 \ifnum #1\expandafter \@firstoftwo
 \else \expandafter \@secondoftwo
 \fi
}%
\providecommand \@ifx [1]{%
 \ifx #1\expandafter \@firstoftwo
 \else \expandafter \@secondoftwo
 \fi
}%
\providecommand \natexlab [1]{#1}%
\providecommand \enquote  [1]{``#1''}%
\providecommand \bibnamefont  [1]{#1}%
\providecommand \bibfnamefont [1]{#1}%
\providecommand \citenamefont [1]{#1}%
\providecommand \href@noop [0]{\@secondoftwo}%
\providecommand \href [0]{\begingroup \@sanitize@url \@href}%
\providecommand \@href[1]{\@@startlink{#1}\@@href}%
\providecommand \@@href[1]{\endgroup#1\@@endlink}%
\providecommand \@sanitize@url [0]{\catcode `\\12\catcode `\$12\catcode
  `\&12\catcode `\#12\catcode `\^12\catcode `\_12\catcode `\%12\relax}%
\providecommand \@@startlink[1]{}%
\providecommand \@@endlink[0]{}%
\providecommand \url  [0]{\begingroup\@sanitize@url \@url }%
\providecommand \@url [1]{\endgroup\@href {#1}{\urlprefix }}%
\providecommand \urlprefix  [0]{URL }%
\providecommand \Eprint [0]{\href }%
\providecommand \doibase [0]{http://dx.doi.org/}%
\providecommand \selectlanguage [0]{\@gobble}%
\providecommand \bibinfo  [0]{\@secondoftwo}%
\providecommand \bibfield  [0]{\@secondoftwo}%
\providecommand \translation [1]{[#1]}%
\providecommand \BibitemOpen [0]{}%
\providecommand \bibitemStop [0]{}%
\providecommand \bibitemNoStop [0]{.\EOS\space}%
\providecommand \EOS [0]{\spacefactor3000\relax}%
\providecommand \BibitemShut  [1]{\csname bibitem#1\endcsname}%
\let\auto@bib@innerbib\@empty
\bibitem [{\citenamefont {D'Ariano}\ \emph {et~al.}(2002)\citenamefont
  {D'Ariano}, \citenamefont {De~Laurentis}, \citenamefont {Paris},
  \citenamefont {Porzio},\ and\ \citenamefont {Solimeno}}]{d2002quantum}%
  \BibitemOpen
  \bibfield  {author} {\bibinfo {author} {\bibfnamefont {G.~M.}\ \bibnamefont
  {D'Ariano}}, \bibinfo {author} {\bibfnamefont {M.}~\bibnamefont
  {De~Laurentis}}, \bibinfo {author} {\bibfnamefont {M.~G.}\ \bibnamefont
  {Paris}}, \bibinfo {author} {\bibfnamefont {A.}~\bibnamefont {Porzio}}, \
  and\ \bibinfo {author} {\bibfnamefont {S.}~\bibnamefont {Solimeno}},\
  }\href@noop {} {\bibfield  {journal} {\bibinfo  {journal} {J. Opt. B: Quantum
  Semiclass. Opt.}\ }\textbf {\bibinfo {volume} {4}},\ \bibinfo {pages} {S127}
  (\bibinfo {year} {2002})}\BibitemShut {NoStop}%
\bibitem [{\citenamefont {H{\"a}ffner}\ \emph {et~al.}(2005)\citenamefont
  {H{\"a}ffner}, \citenamefont {H{\"a}nsel}, \citenamefont {Roos},
  \citenamefont {Benhelm}, \citenamefont {Chwalla}, \citenamefont {K{\"o}rber},
  \citenamefont {Rapol}, \citenamefont {Riebe}, \citenamefont {Schmidt},
  \citenamefont {Becher} \emph {et~al.}}]{haffner2005scalable}%
  \BibitemOpen
  \bibfield  {author} {\bibinfo {author} {\bibfnamefont {H.}~\bibnamefont
  {H{\"a}ffner}}, \bibinfo {author} {\bibfnamefont {W.}~\bibnamefont
  {H{\"a}nsel}}, \bibinfo {author} {\bibfnamefont {C.}~\bibnamefont {Roos}},
  \bibinfo {author} {\bibfnamefont {J.}~\bibnamefont {Benhelm}}, \bibinfo
  {author} {\bibfnamefont {M.}~\bibnamefont {Chwalla}}, \bibinfo {author}
  {\bibfnamefont {T.}~\bibnamefont {K{\"o}rber}}, \bibinfo {author}
  {\bibfnamefont {U.}~\bibnamefont {Rapol}}, \bibinfo {author} {\bibfnamefont
  {M.}~\bibnamefont {Riebe}}, \bibinfo {author} {\bibfnamefont
  {P.}~\bibnamefont {Schmidt}}, \bibinfo {author} {\bibfnamefont
  {C.}~\bibnamefont {Becher}},  \emph {et~al.},\ }\href@noop {} {\bibfield
  {journal} {\bibinfo  {journal} {Nature}\ }\textbf {\bibinfo {volume} {438}},\
  \bibinfo {pages} {643} (\bibinfo {year} {2005})}\BibitemShut {NoStop}%
\bibitem [{\citenamefont {Leibfried}\ \emph {et~al.}(2005)\citenamefont
  {Leibfried}, \citenamefont {Knill}, \citenamefont {Seidelin}, \citenamefont
  {Britton}, \citenamefont {Blakestad}, \citenamefont {Chiaverini},
  \citenamefont {Hume}, \citenamefont {Itano}, \citenamefont {Jost},
  \citenamefont {Langer} \emph {et~al.}}]{leibfried2005creation}%
  \BibitemOpen
  \bibfield  {author} {\bibinfo {author} {\bibfnamefont {D.}~\bibnamefont
  {Leibfried}}, \bibinfo {author} {\bibfnamefont {E.}~\bibnamefont {Knill}},
  \bibinfo {author} {\bibfnamefont {S.}~\bibnamefont {Seidelin}}, \bibinfo
  {author} {\bibfnamefont {J.}~\bibnamefont {Britton}}, \bibinfo {author}
  {\bibfnamefont {R.~B.}\ \bibnamefont {Blakestad}}, \bibinfo {author}
  {\bibfnamefont {J.}~\bibnamefont {Chiaverini}}, \bibinfo {author}
  {\bibfnamefont {D.~B.}\ \bibnamefont {Hume}}, \bibinfo {author}
  {\bibfnamefont {W.~M.}\ \bibnamefont {Itano}}, \bibinfo {author}
  {\bibfnamefont {J.~D.}\ \bibnamefont {Jost}}, \bibinfo {author}
  {\bibfnamefont {C.}~\bibnamefont {Langer}},  \emph {et~al.},\ }\href@noop {}
  {\bibfield  {journal} {\bibinfo  {journal} {Nature}\ }\textbf {\bibinfo
  {volume} {438}},\ \bibinfo {pages} {639} (\bibinfo {year}
  {2005})}\BibitemShut {NoStop}%
\bibitem [{\citenamefont {Lvovsky}\ and\ \citenamefont
  {Raymer}(2009)}]{lvovsky2009continuous}%
  \BibitemOpen
  \bibfield  {author} {\bibinfo {author} {\bibfnamefont {A.~I.}\ \bibnamefont
  {Lvovsky}}\ and\ \bibinfo {author} {\bibfnamefont {M.~G.}\ \bibnamefont
  {Raymer}},\ }\href@noop {} {\bibfield  {journal} {\bibinfo  {journal} {Rev.
  Mod. Phys.}\ }\textbf {\bibinfo {volume} {81}},\ \bibinfo {pages} {299}
  (\bibinfo {year} {2009})}\BibitemShut {NoStop}%
\bibitem [{\citenamefont {Baur}\ \emph {et~al.}(2012)\citenamefont {Baur},
  \citenamefont {Fedorov}, \citenamefont {Steffen}, \citenamefont {Filipp},
  \citenamefont {da~Silva},\ and\ \citenamefont
  {Wallraff}}]{baur2012benchmarking}%
  \BibitemOpen
  \bibfield  {author} {\bibinfo {author} {\bibfnamefont {M.}~\bibnamefont
  {Baur}}, \bibinfo {author} {\bibfnamefont {A.}~\bibnamefont {Fedorov}},
  \bibinfo {author} {\bibfnamefont {L.}~\bibnamefont {Steffen}}, \bibinfo
  {author} {\bibfnamefont {S.}~\bibnamefont {Filipp}}, \bibinfo {author}
  {\bibfnamefont {M.}~\bibnamefont {da~Silva}}, \ and\ \bibinfo {author}
  {\bibfnamefont {A.}~\bibnamefont {Wallraff}},\ }\href@noop {} {\bibfield
  {journal} {\bibinfo  {journal} {Phys. Rev. Lett.}\ }\textbf {\bibinfo
  {volume} {108}},\ \bibinfo {pages} {040502} (\bibinfo {year}
  {2012})}\BibitemShut {NoStop}%
\bibitem [{\citenamefont {Kosaka}\ \emph {et~al.}(2009)\citenamefont {Kosaka},
  \citenamefont {Inagaki}, \citenamefont {Rikitake}, \citenamefont {Imamura},
  \citenamefont {Mitsumori},\ and\ \citenamefont {Edamatsu}}]{KIRI08}%
  \BibitemOpen
  \bibfield  {author} {\bibinfo {author} {\bibfnamefont {H.}~\bibnamefont
  {Kosaka}}, \bibinfo {author} {\bibfnamefont {T.}~\bibnamefont {Inagaki}},
  \bibinfo {author} {\bibfnamefont {Y.}~\bibnamefont {Rikitake}}, \bibinfo
  {author} {\bibfnamefont {H.}~\bibnamefont {Imamura}}, \bibinfo {author}
  {\bibfnamefont {Y.}~\bibnamefont {Mitsumori}}, \ and\ \bibinfo {author}
  {\bibfnamefont {K.}~\bibnamefont {Edamatsu}},\ }\href@noop {} {\bibfield
  {journal} {\bibinfo  {journal} {Nature}\ }\textbf {\bibinfo {volume} {457}},\
  \bibinfo {pages} {702} (\bibinfo {year} {2009})}\BibitemShut {NoStop}%
\bibitem [{\citenamefont {Vanner}\ \emph {et~al.}(2013)\citenamefont {Vanner},
  \citenamefont {Hofer}, \citenamefont {Cole},\ and\ \citenamefont
  {Aspelmeyer}}]{VHCA13}%
  \BibitemOpen
  \bibfield  {author} {\bibinfo {author} {\bibfnamefont {M.}~\bibnamefont
  {Vanner}}, \bibinfo {author} {\bibfnamefont {J.}~\bibnamefont {Hofer}},
  \bibinfo {author} {\bibfnamefont {G.}~\bibnamefont {Cole}}, \ and\ \bibinfo
  {author} {\bibfnamefont {M.}~\bibnamefont {Aspelmeyer}},\ }\href@noop {}
  {\bibfield  {journal} {\bibinfo  {journal} {Nat. Commun.}\ }\textbf {\bibinfo
  {volume} {4}} (\bibinfo {year} {2013})}\BibitemShut {NoStop}%
\bibitem [{\citenamefont {Flammia}\ \emph {et~al.}(2012)\citenamefont
  {Flammia}, \citenamefont {Gross}, \citenamefont {Liu},\ and\ \citenamefont
  {Eisert}}]{FGLE12}%
  \BibitemOpen
  \bibfield  {author} {\bibinfo {author} {\bibfnamefont {S.~T.}\ \bibnamefont
  {Flammia}}, \bibinfo {author} {\bibfnamefont {D.}~\bibnamefont {Gross}},
  \bibinfo {author} {\bibfnamefont {Y.-K.}\ \bibnamefont {Liu}}, \ and\
  \bibinfo {author} {\bibfnamefont {J.}~\bibnamefont {Eisert}},\ }\href@noop {}
  {\bibfield  {journal} {\bibinfo  {journal} {New J. Phys.}\ }\textbf {\bibinfo
  {volume} {14}},\ \bibinfo {pages} {095022} (\bibinfo {year}
  {2012})}\BibitemShut {NoStop}%
\bibitem [{\citenamefont {Gross}\ \emph {et~al.}(2010)\citenamefont {Gross},
  \citenamefont {Liu}, \citenamefont {Flammia}, \citenamefont {Becker},\ and\
  \citenamefont {Eisert}}]{GLFB10}%
  \BibitemOpen
  \bibfield  {author} {\bibinfo {author} {\bibfnamefont {D.}~\bibnamefont
  {Gross}}, \bibinfo {author} {\bibfnamefont {Y.-K.}\ \bibnamefont {Liu}},
  \bibinfo {author} {\bibfnamefont {S.~T.}\ \bibnamefont {Flammia}}, \bibinfo
  {author} {\bibfnamefont {S.}~\bibnamefont {Becker}}, \ and\ \bibinfo {author}
  {\bibfnamefont {J.}~\bibnamefont {Eisert}},\ }\href@noop {} {\bibfield
  {journal} {\bibinfo  {journal} {Phys. Rev. Lett.}\ }\textbf {\bibinfo
  {volume} {105}},\ \bibinfo {pages} {150401} (\bibinfo {year}
  {2010})}\BibitemShut {NoStop}%
\bibitem [{\citenamefont {Lu}\ \emph {et~al.}(2015{\natexlab{a}})\citenamefont
  {Lu}, \citenamefont {Xin}, \citenamefont {Yu}, \citenamefont {Ji},
  \citenamefont {Chen}, \citenamefont {Long}, \citenamefont {Baugh},
  \citenamefont {Peng}, \citenamefont {Zeng},\ and\ \citenamefont
  {Laflamme}}]{LXYJ15}%
  \BibitemOpen
  \bibfield  {author} {\bibinfo {author} {\bibfnamefont {D.}~\bibnamefont
  {Lu}}, \bibinfo {author} {\bibfnamefont {T.}~\bibnamefont {Xin}}, \bibinfo
  {author} {\bibfnamefont {N.}~\bibnamefont {Yu}}, \bibinfo {author}
  {\bibfnamefont {Z.}~\bibnamefont {Ji}}, \bibinfo {author} {\bibfnamefont
  {J.}~\bibnamefont {Chen}}, \bibinfo {author} {\bibfnamefont {G.}~\bibnamefont
  {Long}}, \bibinfo {author} {\bibfnamefont {J.}~\bibnamefont {Baugh}},
  \bibinfo {author} {\bibfnamefont {X.}~\bibnamefont {Peng}}, \bibinfo {author}
  {\bibfnamefont {B.}~\bibnamefont {Zeng}}, \ and\ \bibinfo {author}
  {\bibfnamefont {R.}~\bibnamefont {Laflamme}},\ }\href@noop {} {\bibfield
  {journal} {\bibinfo  {journal} {arXiv:1511.00581}\ } (\bibinfo {year}
  {2015}{\natexlab{a}})}\BibitemShut {NoStop}%
\bibitem [{\citenamefont {Baldwin}\ \emph {et~al.}(2015)\citenamefont
  {Baldwin}, \citenamefont {Deutsch},\ and\ \citenamefont {Kalev}}]{BDK15}%
  \BibitemOpen
  \bibfield  {author} {\bibinfo {author} {\bibfnamefont {C.~H.}\ \bibnamefont
  {Baldwin}}, \bibinfo {author} {\bibfnamefont {I.~H.}\ \bibnamefont
  {Deutsch}}, \ and\ \bibinfo {author} {\bibfnamefont {A.}~\bibnamefont
  {Kalev}},\ }\href@noop {} {\bibfield  {journal} {\bibinfo  {journal}
  {arXiv:1510.02736}\ } (\bibinfo {year} {2015})}\BibitemShut {NoStop}%
\bibitem [{\citenamefont {Kalev}\ and\ \citenamefont {Baldwin}(2015)}]{KB15}%
  \BibitemOpen
  \bibfield  {author} {\bibinfo {author} {\bibfnamefont {A.}~\bibnamefont
  {Kalev}}\ and\ \bibinfo {author} {\bibfnamefont {C.~H.}\ \bibnamefont
  {Baldwin}},\ }\href@noop {} {\bibfield  {journal} {\bibinfo  {journal}
  {arXiv:1511.01433}\ } (\bibinfo {year} {2015})}\BibitemShut {NoStop}%
\bibitem [{\citenamefont {Haah}\ \emph {et~al.}(2015)\citenamefont {Haah},
  \citenamefont {Harrow}, \citenamefont {Ji}, \citenamefont {Wu},\ and\
  \citenamefont {Yu}}]{HHJW16}%
  \BibitemOpen
  \bibfield  {author} {\bibinfo {author} {\bibfnamefont {J.}~\bibnamefont
  {Haah}}, \bibinfo {author} {\bibfnamefont {A.~W.}\ \bibnamefont {Harrow}},
  \bibinfo {author} {\bibfnamefont {Z.}~\bibnamefont {Ji}}, \bibinfo {author}
  {\bibfnamefont {X.}~\bibnamefont {Wu}}, \ and\ \bibinfo {author}
  {\bibfnamefont {N.}~\bibnamefont {Yu}},\ }\href@noop {} {\bibfield  {journal}
  {\bibinfo  {journal} {arXiv:1508.01797}\ } (\bibinfo {year}
  {2015})}\BibitemShut {NoStop}%
\bibitem [{\citenamefont {Linden}\ \emph {et~al.}(2002)\citenamefont {Linden},
  \citenamefont {Popescu},\ and\ \citenamefont {Wootters}}]{Linden2002}%
  \BibitemOpen
  \bibfield  {author} {\bibinfo {author} {\bibfnamefont {N.}~\bibnamefont
  {Linden}}, \bibinfo {author} {\bibfnamefont {S.}~\bibnamefont {Popescu}}, \
  and\ \bibinfo {author} {\bibfnamefont {W.~K.}\ \bibnamefont {Wootters}},\
  }\href@noop {} {\bibfield  {journal} {\bibinfo  {journal} {Phys. Rev. Lett.}\
  }\textbf {\bibinfo {volume} {89}},\ \bibinfo {pages} {207901} (\bibinfo
  {year} {2002})}\BibitemShut {NoStop}%
\bibitem [{\citenamefont {Linden}\ and\ \citenamefont
  {Wootters}(2002)}]{Linden2002b}%
  \BibitemOpen
  \bibfield  {author} {\bibinfo {author} {\bibfnamefont {N.}~\bibnamefont
  {Linden}}\ and\ \bibinfo {author} {\bibfnamefont {W.~K.}\ \bibnamefont
  {Wootters}},\ }\href@noop {} {\bibfield  {journal} {\bibinfo  {journal}
  {Phys. Rev. Lett.}\ }\textbf {\bibinfo {volume} {89}},\ \bibinfo {pages}
  {277906} (\bibinfo {year} {2002})}\BibitemShut {NoStop}%
\bibitem [{\citenamefont {Di\'osi}(2004)}]{Diosi2004}%
  \BibitemOpen
  \bibfield  {author} {\bibinfo {author} {\bibfnamefont {L.}~\bibnamefont
  {Di\'osi}},\ }\href@noop {} {\bibfield  {journal} {\bibinfo  {journal} {Phys.
  Rev. A}\ }\textbf {\bibinfo {volume} {70}},\ \bibinfo {pages} {010302}
  (\bibinfo {year} {2004})}\BibitemShut {NoStop}%
\bibitem [{\citenamefont {Chen}\ \emph
  {et~al.}(2012{\natexlab{a}})\citenamefont {Chen}, \citenamefont {Ji},
  \citenamefont {Ruskai}, \citenamefont {Zeng},\ and\ \citenamefont
  {Zhou}}]{CJRZ12}%
  \BibitemOpen
  \bibfield  {author} {\bibinfo {author} {\bibfnamefont {J.}~\bibnamefont
  {Chen}}, \bibinfo {author} {\bibfnamefont {Z.}~\bibnamefont {Ji}}, \bibinfo
  {author} {\bibfnamefont {M.~B.}\ \bibnamefont {Ruskai}}, \bibinfo {author}
  {\bibfnamefont {B.}~\bibnamefont {Zeng}}, \ and\ \bibinfo {author}
  {\bibfnamefont {D.-L.}\ \bibnamefont {Zhou}},\ }\href@noop {} {\bibfield
  {journal} {\bibinfo  {journal} {J. Math. Phys.}\ }\textbf {\bibinfo {volume}
  {53}},\ \bibinfo {pages} {072203} (\bibinfo {year}
  {2012}{\natexlab{a}})}\BibitemShut {NoStop}%
\bibitem [{\citenamefont {Chen}\ \emph
  {et~al.}(2012{\natexlab{b}})\citenamefont {Chen}, \citenamefont {Ji},
  \citenamefont {Zeng},\ and\ \citenamefont {Zhou}}]{Chen2012}%
  \BibitemOpen
  \bibfield  {author} {\bibinfo {author} {\bibfnamefont {J.}~\bibnamefont
  {Chen}}, \bibinfo {author} {\bibfnamefont {Z.}~\bibnamefont {Ji}}, \bibinfo
  {author} {\bibfnamefont {B.}~\bibnamefont {Zeng}}, \ and\ \bibinfo {author}
  {\bibfnamefont {D.~L.}\ \bibnamefont {Zhou}},\ }\href@noop {} {\bibfield
  {journal} {\bibinfo  {journal} {Phys. Rev. A}\ }\textbf {\bibinfo {volume}
  {86}},\ \bibinfo {pages} {022339} (\bibinfo {year}
  {2012}{\natexlab{b}})}\BibitemShut {NoStop}%
\bibitem [{\citenamefont {Chen}\ \emph {et~al.}(2013)\citenamefont {Chen},
  \citenamefont {Dawkins}, \citenamefont {Ji}, \citenamefont {Johnston},
  \citenamefont {Kribs}, \citenamefont {Shultz},\ and\ \citenamefont
  {Zeng}}]{Chen2013}%
  \BibitemOpen
  \bibfield  {author} {\bibinfo {author} {\bibfnamefont {J.}~\bibnamefont
  {Chen}}, \bibinfo {author} {\bibfnamefont {H.}~\bibnamefont {Dawkins}},
  \bibinfo {author} {\bibfnamefont {Z.}~\bibnamefont {Ji}}, \bibinfo {author}
  {\bibfnamefont {N.}~\bibnamefont {Johnston}}, \bibinfo {author}
  {\bibfnamefont {D.}~\bibnamefont {Kribs}}, \bibinfo {author} {\bibfnamefont
  {F.}~\bibnamefont {Shultz}}, \ and\ \bibinfo {author} {\bibfnamefont
  {B.}~\bibnamefont {Zeng}},\ }\href@noop {} {\bibfield  {journal} {\bibinfo
  {journal} {Phys. Rev. A}\ }\textbf {\bibinfo {volume} {88}},\ \bibinfo
  {pages} {012109} (\bibinfo {year} {2013})}\BibitemShut {NoStop}%
\bibitem [{sup()}]{supple}%
  \BibitemOpen
  \href@noop {} {\bibinfo  {journal} {See Appendix for details.}\ }\BibitemShut
  {NoStop}%
\bibitem [{\citenamefont {Hastings}(2010)}]{hastings2010locality}%
  \BibitemOpen
\bibfield  {journal} {  }\bibfield  {author} {\bibinfo {author} {\bibfnamefont
  {M.~B.}\ \bibnamefont {Hastings}},\ }\href@noop {} {\bibfield  {journal}
  {\bibinfo  {journal} {arXiv:1008.5137}\ } (\bibinfo {year}
  {2010})}\BibitemShut {NoStop}%
\bibitem [{\citenamefont {Wolf}\ \emph {et~al.}(2008)\citenamefont {Wolf},
  \citenamefont {Verstraete}, \citenamefont {Hastings},\ and\ \citenamefont
  {Cirac}}]{wolf2008area}%
  \BibitemOpen
  \bibfield  {author} {\bibinfo {author} {\bibfnamefont {M.~M.}\ \bibnamefont
  {Wolf}}, \bibinfo {author} {\bibfnamefont {F.}~\bibnamefont {Verstraete}},
  \bibinfo {author} {\bibfnamefont {M.~B.}\ \bibnamefont {Hastings}}, \ and\
  \bibinfo {author} {\bibfnamefont {J.~I.}\ \bibnamefont {Cirac}},\ }\href@noop
  {} {\bibfield  {journal} {\bibinfo  {journal} {Phys. Rev. Lett.}\ }\textbf
  {\bibinfo {volume} {100}},\ \bibinfo {pages} {070502} (\bibinfo {year}
  {2008})}\BibitemShut {NoStop}%
\bibitem [{\citenamefont {Perez-Garcia}\ \emph {et~al.}(2006)\citenamefont
  {Perez-Garcia}, \citenamefont {Verstraete}, \citenamefont {Wolf},\ and\
  \citenamefont {Cirac}}]{perez2006matrix}%
  \BibitemOpen
  \bibfield  {author} {\bibinfo {author} {\bibfnamefont {D.}~\bibnamefont
  {Perez-Garcia}}, \bibinfo {author} {\bibfnamefont {F.}~\bibnamefont
  {Verstraete}}, \bibinfo {author} {\bibfnamefont {M.~M.}\ \bibnamefont
  {Wolf}}, \ and\ \bibinfo {author} {\bibfnamefont {J.~I.}\ \bibnamefont
  {Cirac}},\ }\href@noop {} {\bibfield  {journal} {\bibinfo  {journal}
  {arXiv:quant-ph/0608197}\ } (\bibinfo {year} {2006})}\BibitemShut {NoStop}%
\bibitem [{\citenamefont {Verstraete}\ \emph {et~al.}(2008)\citenamefont
  {Verstraete}, \citenamefont {Murg},\ and\ \citenamefont
  {Cirac}}]{verstraete2008matrix}%
  \BibitemOpen
  \bibfield  {author} {\bibinfo {author} {\bibfnamefont {F.}~\bibnamefont
  {Verstraete}}, \bibinfo {author} {\bibfnamefont {V.}~\bibnamefont {Murg}}, \
  and\ \bibinfo {author} {\bibfnamefont {J.~I.}\ \bibnamefont {Cirac}},\
  }\href@noop {} {\bibfield  {journal} {\bibinfo  {journal} {Adv. Phys.}\
  }\textbf {\bibinfo {volume} {57}},\ \bibinfo {pages} {143} (\bibinfo {year}
  {2008})}\BibitemShut {NoStop}%
\bibitem [{\citenamefont {Cirac}\ and\ \citenamefont
  {Verstraete}(2009)}]{cirac2009renormalization}%
  \BibitemOpen
  \bibfield  {author} {\bibinfo {author} {\bibfnamefont {J.~I.}\ \bibnamefont
  {Cirac}}\ and\ \bibinfo {author} {\bibfnamefont {F.}~\bibnamefont
  {Verstraete}},\ }\href@noop {} {\bibfield  {journal} {\bibinfo  {journal} {J.
  Phys. A}\ }\textbf {\bibinfo {volume} {42}},\ \bibinfo {pages} {504004}
  (\bibinfo {year} {2009})}\BibitemShut {NoStop}%
\bibitem [{\citenamefont {Cramer}\ \emph {et~al.}(2010)\citenamefont {Cramer},
  \citenamefont {Plenio}, \citenamefont {Flammia}, \citenamefont {Somma},
  \citenamefont {Gross}, \citenamefont {Bartlett}, \citenamefont
  {Landon-Cardinal}, \citenamefont {Poulin},\ and\ \citenamefont
  {Liu}}]{cramer2010efficient}%
  \BibitemOpen
  \bibfield  {author} {\bibinfo {author} {\bibfnamefont {M.}~\bibnamefont
  {Cramer}}, \bibinfo {author} {\bibfnamefont {M.~B.}\ \bibnamefont {Plenio}},
  \bibinfo {author} {\bibfnamefont {S.~T.}\ \bibnamefont {Flammia}}, \bibinfo
  {author} {\bibfnamefont {R.}~\bibnamefont {Somma}}, \bibinfo {author}
  {\bibfnamefont {D.}~\bibnamefont {Gross}}, \bibinfo {author} {\bibfnamefont
  {S.~D.}\ \bibnamefont {Bartlett}}, \bibinfo {author} {\bibfnamefont
  {O.}~\bibnamefont {Landon-Cardinal}}, \bibinfo {author} {\bibfnamefont
  {D.}~\bibnamefont {Poulin}}, \ and\ \bibinfo {author} {\bibfnamefont {Y.-K.}\
  \bibnamefont {Liu}},\ }\href@noop {} {\bibfield  {journal} {\bibinfo
  {journal} {Nat. Commun.}\ }\textbf {\bibinfo {volume} {1}},\ \bibinfo {pages}
  {149} (\bibinfo {year} {2010})}\BibitemShut {NoStop}%
\bibitem [{\citenamefont {Pauli}(1958)}]{pauli1958allgemeinen}%
  \BibitemOpen
  \bibfield  {author} {\bibinfo {author} {\bibfnamefont {W.}~\bibnamefont
  {Pauli}},\ }\href@noop {} {\emph {\bibinfo {title} {Die allgemeinen
  prinzipien der wellenmechanik}}}\ (\bibinfo  {publisher} {Springer Berlin},\
  \bibinfo {year} {1958})\BibitemShut {NoStop}%
\bibitem [{\citenamefont {Weigert}(1992)}]{weigert1992pauli}%
  \BibitemOpen
  \bibfield  {author} {\bibinfo {author} {\bibfnamefont {S.}~\bibnamefont
  {Weigert}},\ }\href@noop {} {\bibfield  {journal} {\bibinfo  {journal} {Phys.
  Rev. A}\ }\textbf {\bibinfo {volume} {45}},\ \bibinfo {pages} {7688}
  (\bibinfo {year} {1992})}\BibitemShut {NoStop}%
\bibitem [{\citenamefont {Heinosaari}\ \emph {et~al.}(2013)\citenamefont
  {Heinosaari}, \citenamefont {Mazzarella},\ and\ \citenamefont
  {Wolf}}]{heinosaari2013quantum}%
  \BibitemOpen
  \bibfield  {author} {\bibinfo {author} {\bibfnamefont {T.}~\bibnamefont
  {Heinosaari}}, \bibinfo {author} {\bibfnamefont {L.}~\bibnamefont
  {Mazzarella}}, \ and\ \bibinfo {author} {\bibfnamefont {M.~M.}\ \bibnamefont
  {Wolf}},\ }\href@noop {} {\bibfield  {journal} {\bibinfo  {journal} {Commun.
  Math. Phys.}\ }\textbf {\bibinfo {volume} {318}},\ \bibinfo {pages} {355}
  (\bibinfo {year} {2013})}\BibitemShut {NoStop}%
\bibitem [{\citenamefont {Eckert}\ \emph {et~al.}(2002)\citenamefont {Eckert},
  \citenamefont {Schliemann}, \citenamefont {Bruss},\ and\ \citenamefont
  {Lewenstein}}]{ESBL02}%
  \BibitemOpen
  \bibfield  {author} {\bibinfo {author} {\bibfnamefont {K.}~\bibnamefont
  {Eckert}}, \bibinfo {author} {\bibfnamefont {J.}~\bibnamefont {Schliemann}},
  \bibinfo {author} {\bibfnamefont {D.}~\bibnamefont {Bruss}}, \ and\ \bibinfo
  {author} {\bibfnamefont {M.}~\bibnamefont {Lewenstein}},\ }\href@noop {}
  {\bibfield  {journal} {\bibinfo  {journal} {Annals of Physics}\ }\textbf
  {\bibinfo {volume} {299}},\ \bibinfo {pages} {88} (\bibinfo {year}
  {2002})}\BibitemShut {NoStop}%
\bibitem [{\citenamefont {Bastin}\ \emph {et~al.}(2009)\citenamefont {Bastin},
  \citenamefont {Krins}, \citenamefont {Mathonet}, \citenamefont {Godefroid},
  \citenamefont {Lamata},\ and\ \citenamefont {Solano}}]{BKMG+09}%
  \BibitemOpen
  \bibfield  {author} {\bibinfo {author} {\bibfnamefont {T.}~\bibnamefont
  {Bastin}}, \bibinfo {author} {\bibfnamefont {S.}~\bibnamefont {Krins}},
  \bibinfo {author} {\bibfnamefont {P.}~\bibnamefont {Mathonet}}, \bibinfo
  {author} {\bibfnamefont {M.}~\bibnamefont {Godefroid}}, \bibinfo {author}
  {\bibfnamefont {L.}~\bibnamefont {Lamata}}, \ and\ \bibinfo {author}
  {\bibfnamefont {E.}~\bibnamefont {Solano}},\ }\href@noop {} {\bibfield
  {journal} {\bibinfo  {journal} {Phys. Rev. Lett.}\ }\textbf {\bibinfo
  {volume} {103}},\ \bibinfo {pages} {070503} (\bibinfo {year}
  {2009})}\BibitemShut {NoStop}%
\bibitem [{\citenamefont {Yu}\ \emph {et~al.}(2014)\citenamefont {Yu},
  \citenamefont {Guo},\ and\ \citenamefont {Duan}}]{YU14}%
  \BibitemOpen
  \bibfield  {author} {\bibinfo {author} {\bibfnamefont {N.}~\bibnamefont
  {Yu}}, \bibinfo {author} {\bibfnamefont {C.}~\bibnamefont {Guo}}, \ and\
  \bibinfo {author} {\bibfnamefont {R.}~\bibnamefont {Duan}},\ }\href@noop {}
  {\bibfield  {journal} {\bibinfo  {journal} {Phys. Rev. Lett.}\ }\textbf
  {\bibinfo {volume} {112}},\ \bibinfo {pages} {160401} (\bibinfo {year}
  {2014})}\BibitemShut {NoStop}%
\bibitem [{\citenamefont {Cramer}\ \emph {et~al.}(2013)\citenamefont {Cramer},
  \citenamefont {Bernard}, \citenamefont {Fabbri}, \citenamefont {Fallani},
  \citenamefont {Fort}, \citenamefont {Rosi}, \citenamefont {Caruso},
  \citenamefont {Inguscio},\ and\ \citenamefont {Plenio}}]{CBF+13}%
  \BibitemOpen
  \bibfield  {author} {\bibinfo {author} {\bibfnamefont {M.}~\bibnamefont
  {Cramer}}, \bibinfo {author} {\bibfnamefont {A.}~\bibnamefont {Bernard}},
  \bibinfo {author} {\bibfnamefont {N.}~\bibnamefont {Fabbri}}, \bibinfo
  {author} {\bibfnamefont {L.}~\bibnamefont {Fallani}}, \bibinfo {author}
  {\bibfnamefont {C.}~\bibnamefont {Fort}}, \bibinfo {author} {\bibfnamefont
  {S.}~\bibnamefont {Rosi}}, \bibinfo {author} {\bibfnamefont {F.}~\bibnamefont
  {Caruso}}, \bibinfo {author} {\bibfnamefont {M.}~\bibnamefont {Inguscio}}, \
  and\ \bibinfo {author} {\bibfnamefont {M.}~\bibnamefont {Plenio}},\
  }\href@noop {} {\bibfield  {journal} {\bibinfo  {journal} {Nat. Commun.}\
  }\textbf {\bibinfo {volume} {4}},\ \bibinfo {pages} {2161} (\bibinfo {year}
  {2013})}\BibitemShut {NoStop}%
\bibitem [{\citenamefont {McConnell}\ \emph {et~al.}(2015)\citenamefont
  {McConnell}, \citenamefont {Zhang}, \citenamefont {Hu}, \citenamefont {Cuk},\
  and\ \citenamefont {Vuletic}}]{MZHC15}%
  \BibitemOpen
  \bibfield  {author} {\bibinfo {author} {\bibfnamefont {R.}~\bibnamefont
  {McConnell}}, \bibinfo {author} {\bibfnamefont {H.}~\bibnamefont {Zhang}},
  \bibinfo {author} {\bibfnamefont {J.}~\bibnamefont {Hu}}, \bibinfo {author}
  {\bibfnamefont {S.}~\bibnamefont {Cuk}}, \ and\ \bibinfo {author}
  {\bibfnamefont {V.}~\bibnamefont {Vuletic}},\ }\href@noop {} {\bibfield
  {journal} {\bibinfo  {journal} {Nature}\ }\textbf {\bibinfo {volume} {519}},\
  \bibinfo {pages} {439} (\bibinfo {year} {2015})}\BibitemShut {NoStop}%
\bibitem [{\citenamefont {Parashar}\ and\ \citenamefont
  {Rana}(2009)}]{parashar2009n}%
  \BibitemOpen
  \bibfield  {author} {\bibinfo {author} {\bibfnamefont {P.}~\bibnamefont
  {Parashar}}\ and\ \bibinfo {author} {\bibfnamefont {S.}~\bibnamefont
  {Rana}},\ }\href@noop {} {\bibfield  {journal} {\bibinfo  {journal} {Phys.
  Rev. A}\ }\textbf {\bibinfo {volume} {80}},\ \bibinfo {pages} {012319}
  (\bibinfo {year} {2009})}\BibitemShut {NoStop}%
\bibitem [{\citenamefont {Cory}\ \emph {et~al.}(1997)\citenamefont {Cory},
  \citenamefont {Fahmy},\ and\ \citenamefont {Havel}}]{cory1997ensemble}%
  \BibitemOpen
  \bibfield  {author} {\bibinfo {author} {\bibfnamefont {D.~G.}\ \bibnamefont
  {Cory}}, \bibinfo {author} {\bibfnamefont {A.~F.}\ \bibnamefont {Fahmy}}, \
  and\ \bibinfo {author} {\bibfnamefont {T.~F.}\ \bibnamefont {Havel}},\
  }\href@noop {} {\bibfield  {journal} {\bibinfo  {journal} {Proc. Natl. Acad.
  Sci.}\ }\textbf {\bibinfo {volume} {94}},\ \bibinfo {pages} {1634} (\bibinfo
  {year} {1997})}\BibitemShut {NoStop}%
\bibitem [{\citenamefont {Lu}\ \emph {et~al.}(2011)\citenamefont {Lu},
  \citenamefont {Xu}, \citenamefont {Xu}, \citenamefont {Chen}, \citenamefont
  {Gong}, \citenamefont {Peng},\ and\ \citenamefont {Du}}]{lu2011simulation}%
  \BibitemOpen
  \bibfield  {author} {\bibinfo {author} {\bibfnamefont {D.}~\bibnamefont
  {Lu}}, \bibinfo {author} {\bibfnamefont {N.}~\bibnamefont {Xu}}, \bibinfo
  {author} {\bibfnamefont {R.}~\bibnamefont {Xu}}, \bibinfo {author}
  {\bibfnamefont {H.}~\bibnamefont {Chen}}, \bibinfo {author} {\bibfnamefont
  {J.}~\bibnamefont {Gong}}, \bibinfo {author} {\bibfnamefont {X.}~\bibnamefont
  {Peng}}, \ and\ \bibinfo {author} {\bibfnamefont {J.}~\bibnamefont {Du}},\
  }\href@noop {} {\bibfield  {journal} {\bibinfo  {journal} {Phys. Rev. Lett.}\
  }\textbf {\bibinfo {volume} {107}},\ \bibinfo {pages} {020501} (\bibinfo
  {year} {2011})}\BibitemShut {NoStop}%
\bibitem [{\citenamefont {Xin}\ \emph {et~al.}(2015)\citenamefont {Xin},
  \citenamefont {Li}, \citenamefont {Wang},\ and\ \citenamefont
  {Long}}]{PhysRevA.92.022126}%
  \BibitemOpen
  \bibfield  {author} {\bibinfo {author} {\bibfnamefont {T.}~\bibnamefont
  {Xin}}, \bibinfo {author} {\bibfnamefont {H.}~\bibnamefont {Li}}, \bibinfo
  {author} {\bibfnamefont {B.-X.}\ \bibnamefont {Wang}}, \ and\ \bibinfo
  {author} {\bibfnamefont {G.-L.}\ \bibnamefont {Long}},\ }\href@noop {}
  {\bibfield  {journal} {\bibinfo  {journal} {Phys. Rev. A}\ }\textbf {\bibinfo
  {volume} {92}},\ \bibinfo {pages} {022126} (\bibinfo {year}
  {2015})}\BibitemShut {NoStop}%
\bibitem [{\citenamefont {Khaneja}\ \emph {et~al.}(2005)\citenamefont
  {Khaneja}, \citenamefont {Reiss}, \citenamefont {Kehlet}, \citenamefont
  {Schulte-Herbr{\"u}ggen},\ and\ \citenamefont {Glaser}}]{khaneja2005optimal}%
  \BibitemOpen
  \bibfield  {author} {\bibinfo {author} {\bibfnamefont {N.}~\bibnamefont
  {Khaneja}}, \bibinfo {author} {\bibfnamefont {T.}~\bibnamefont {Reiss}},
  \bibinfo {author} {\bibfnamefont {C.}~\bibnamefont {Kehlet}}, \bibinfo
  {author} {\bibfnamefont {T.}~\bibnamefont {Schulte-Herbr{\"u}ggen}}, \ and\
  \bibinfo {author} {\bibfnamefont {S.~J.}\ \bibnamefont {Glaser}},\
  }\href@noop {} {\bibfield  {journal} {\bibinfo  {journal} {J. Magn. Reson.}\
  }\textbf {\bibinfo {volume} {172}},\ \bibinfo {pages} {296} (\bibinfo {year}
  {2005})}\BibitemShut {NoStop}%
\bibitem [{\citenamefont {Ryan}\ \emph {et~al.}(2008)\citenamefont {Ryan},
  \citenamefont {Negrevergne}, \citenamefont {Laforest}, \citenamefont
  {Knill},\ and\ \citenamefont {Laflamme}}]{ryan2008liquid}%
  \BibitemOpen
  \bibfield  {author} {\bibinfo {author} {\bibfnamefont {C.}~\bibnamefont
  {Ryan}}, \bibinfo {author} {\bibfnamefont {C.}~\bibnamefont {Negrevergne}},
  \bibinfo {author} {\bibfnamefont {M.}~\bibnamefont {Laforest}}, \bibinfo
  {author} {\bibfnamefont {E.}~\bibnamefont {Knill}}, \ and\ \bibinfo {author}
  {\bibfnamefont {R.}~\bibnamefont {Laflamme}},\ }\href@noop {} {\bibfield
  {journal} {\bibinfo  {journal} {Phys. Rev. A}\ }\textbf {\bibinfo {volume}
  {78}},\ \bibinfo {pages} {012328} (\bibinfo {year} {2008})}\BibitemShut
  {NoStop}%
\bibitem [{\citenamefont {Moussa}\ \emph {et~al.}(2012)\citenamefont {Moussa},
  \citenamefont {da~Silva}, \citenamefont {Ryan},\ and\ \citenamefont
  {Laflamme}}]{moussa2012practical}%
  \BibitemOpen
  \bibfield  {author} {\bibinfo {author} {\bibfnamefont {O.}~\bibnamefont
  {Moussa}}, \bibinfo {author} {\bibfnamefont {M.~P.}\ \bibnamefont
  {da~Silva}}, \bibinfo {author} {\bibfnamefont {C.~A.}\ \bibnamefont {Ryan}},
  \ and\ \bibinfo {author} {\bibfnamefont {R.}~\bibnamefont {Laflamme}},\
  }\href@noop {} {\bibfield  {journal} {\bibinfo  {journal} {Phys. Rev. Lett.}\
  }\textbf {\bibinfo {volume} {109}},\ \bibinfo {pages} {070504} (\bibinfo
  {year} {2012})}\BibitemShut {NoStop}%
\bibitem [{\citenamefont {Lu}\ \emph {et~al.}(2015{\natexlab{b}})\citenamefont
  {Lu}, \citenamefont {Li}, \citenamefont {Trottier}, \citenamefont {Li},
  \citenamefont {Brodutch}, \citenamefont {Krismanich}, \citenamefont
  {Ghavami}, \citenamefont {Dmitrienko}, \citenamefont {Long}, \citenamefont
  {Baugh} \emph {et~al.}}]{lu2015experimental}%
  \BibitemOpen
  \bibfield  {author} {\bibinfo {author} {\bibfnamefont {D.}~\bibnamefont
  {Lu}}, \bibinfo {author} {\bibfnamefont {H.}~\bibnamefont {Li}}, \bibinfo
  {author} {\bibfnamefont {D.-A.}\ \bibnamefont {Trottier}}, \bibinfo {author}
  {\bibfnamefont {J.}~\bibnamefont {Li}}, \bibinfo {author} {\bibfnamefont
  {A.}~\bibnamefont {Brodutch}}, \bibinfo {author} {\bibfnamefont {A.~P.}\
  \bibnamefont {Krismanich}}, \bibinfo {author} {\bibfnamefont
  {A.}~\bibnamefont {Ghavami}}, \bibinfo {author} {\bibfnamefont {G.~I.}\
  \bibnamefont {Dmitrienko}}, \bibinfo {author} {\bibfnamefont
  {G.}~\bibnamefont {Long}}, \bibinfo {author} {\bibfnamefont {J.}~\bibnamefont
  {Baugh}}, \ and\ \bibinfo {author} {\bibfnamefont {R.}~\bibnamefont
  {Laflamme}},\ }\href@noop {} {\bibfield  {journal} {\bibinfo
  {journal} {Phys. Rev. Lett.}\ }\textbf {\bibinfo {volume} {114}},\ \bibinfo
  {pages} {140505} (\bibinfo {year} {2015}{\natexlab{b}})}\BibitemShut
  {NoStop}%
\bibitem [{\citenamefont {Leskowitz}\ and\ \citenamefont
  {Mueller}(2004)}]{leskowitz2004state}%
  \BibitemOpen
  \bibfield  {author} {\bibinfo {author} {\bibfnamefont {G.~M.}\ \bibnamefont
  {Leskowitz}}\ and\ \bibinfo {author} {\bibfnamefont {L.~J.}\ \bibnamefont
  {Mueller}},\ }\href@noop {} {\bibfield  {journal} {\bibinfo  {journal} {Phys.
  Rev. A}\ }\textbf {\bibinfo {volume} {69}},\ \bibinfo {pages} {052302}
  (\bibinfo {year} {2004})}\BibitemShut {NoStop}%
\bibitem [{\citenamefont {Lee}(2002)}]{lee2002quantum}%
  \BibitemOpen
  \bibfield  {author} {\bibinfo {author} {\bibfnamefont {J.-S.}\ \bibnamefont
  {Lee}},\ }\href@noop {} {\bibfield  {journal} {\bibinfo  {journal} {Phys.
  Rev. A}\ }\textbf {\bibinfo {volume} {305}},\ \bibinfo {pages} {349}
  (\bibinfo {year} {2002})}\BibitemShut {NoStop}%
\bibitem [{\citenamefont {Altepeter}\ \emph {et~al.}(2005)\citenamefont
  {Altepeter}, \citenamefont {Jeffrey},\ and\ \citenamefont
  {Kwiat}}]{altepeter2005photonic}%
  \BibitemOpen
  \bibfield  {author} {\bibinfo {author} {\bibfnamefont {J.~B.}\ \bibnamefont
  {Altepeter}}, \bibinfo {author} {\bibfnamefont {E.~R.}\ \bibnamefont
  {Jeffrey}}, \ and\ \bibinfo {author} {\bibfnamefont {P.~G.}\ \bibnamefont
  {Kwiat}},\ }\href@noop {} {\bibfield  {journal} {\bibinfo  {journal} {Adv.
  At. Mol. Opt. Phy.}\ }\textbf {\bibinfo {volume} {52}},\ \bibinfo {pages}
  {105} (\bibinfo {year} {2005})}\BibitemShut {NoStop}%
\bibitem [{\citenamefont {Jones}\ and\ \citenamefont {Linden}(2005)}]{JL05}%
  \BibitemOpen
  \bibfield  {author} {\bibinfo {author} {\bibfnamefont {N.~S.}\ \bibnamefont
  {Jones}}\ and\ \bibinfo {author} {\bibfnamefont {N.}~\bibnamefont {Linden}},\
  }\href {\doibase 10.1103/PhysRevA.71.012324} {\bibfield  {journal} {\bibinfo
  {journal} {Phys. Rev. A}\ }\textbf {\bibinfo {volume} {71}},\ \bibinfo
  {pages} {012324} (\bibinfo {year} {2005})}\BibitemShut {NoStop}%
\bibitem [{\citenamefont {Horodecki}\ \emph {et~al.}(1996)\citenamefont
  {Horodecki}, \citenamefont {Horodecki},\ and\ \citenamefont
  {Horodecki}}]{Horodecki1996}%
  \BibitemOpen
  \bibfield  {author} {\bibinfo {author} {\bibfnamefont {M.}~\bibnamefont
  {Horodecki}}, \bibinfo {author} {\bibfnamefont {P.}~\bibnamefont
  {Horodecki}}, \ and\ \bibinfo {author} {\bibfnamefont {R.}~\bibnamefont
  {Horodecki}},\ }\href@noop {} {\bibfield  {journal} {\bibinfo  {journal}
  {Phys. Lett. A}\ }\textbf {\bibinfo {volume} {223}},\ \bibinfo {pages} {1}
  (\bibinfo {year} {1996})}\BibitemShut {NoStop}%
\bibitem [{\citenamefont {Leinaas}\ \emph {et~al.}(2006)\citenamefont
  {Leinaas}, \citenamefont {Myrheim},\ and\ \citenamefont
  {Ovrum}}]{Leinaas2006}%
  \BibitemOpen
  \bibfield  {author} {\bibinfo {author} {\bibfnamefont {J.~M.}\ \bibnamefont
  {Leinaas}}, \bibinfo {author} {\bibfnamefont {J.}~\bibnamefont {Myrheim}}, \
  and\ \bibinfo {author} {\bibfnamefont {E.}~\bibnamefont {Ovrum}},\
  }\href@noop {} {\bibfield  {journal} {\bibinfo  {journal} {Phys. Rev. A}\
  }\textbf {\bibinfo {volume} {74}},\ \bibinfo {pages} {012313} (\bibinfo
  {year} {2006})}\BibitemShut {NoStop}%
\end{thebibliography}
\end{document}